\pdfoutput=1
\documentclass[preprint,journal]{vgtc}       

\usepackage[hyphens]{url}

\usepackage{mathptmx}
\usepackage{graphicx}
\usepackage{times}
\usepackage{tabularx}
\usepackage{booktabs}

\usepackage{adjustbox}
\usepackage{verbatim}
\usepackage{enumitem}
\usepackage{nicefrac}
\usepackage{subcaption}
\usepackage{makecell}
\usepackage[linewidth=1pt]{mdframed}
\usepackage{tikz}
\usepackage{multirow}
\usepackage{floatrow}
\usepackage[title]{appendix}
\usepackage[rightcaption]{sidecap}
\sidecaptionvpos{figure}{t}

\definecolor{type}{HTML}{EECA3B}
\definecolor{statistical_q}{HTML}{4C78A8}
\definecolor{statistical_c}{HTML}{F58518}
\definecolor{sequence}{HTML}{54A24B}
\definecolor{space}{HTML}{E45756}
\definecolor{outlier}{HTML}{72B7B2}
\definecolor{pairwise}{HTML}{FF9DA6}
\definecolor{name}{HTML}{b279A2}
\definecolor{dimensions}{HTML}{BAB0AC}
\definecolor{unique}{HTML}{9D755D}

\usepackage{amsfonts}
\usepackage{amsmath}
\DeclareMathOperator*{\argmax}{arg\,max}

\DeclareMathAlphabet{\mathcal}{OMS}{cmsy}{m}{n}

\DeclareRobustCommand\colorlegend[1]{\begin{tikzpicture}\fill[fill={#1}] (11.1,5.5) rectangle ++(0.4,0.2);\end{tikzpicture}}

\usepackage{textcomp}

\def\textyen{{\setbox0=\hbox{Y}Y\kern-.97\wd0\vbox{\hrule height.1ex
width.98\wd0\kern.33ex\hrule height.1ex width.98\wd0\kern.45ex}}}

\usepackage[bookmarks,backref=page,linkcolor=black]{hyperref} 
\hypersetup{
  pdfauthor = {},
  pdftitle = {},
  pdfsubject = {},
  pdfkeywords = {},
  colorlinks=true,
  linkcolor= black,
  citecolor= black,
  pageanchor=true,
  urlcolor = black,
  plainpages = false,
  breaklinks = true,
  linktocpage
}

\usepackage{colortbl}
\usepackage{cleveref}

\crefformat{section}{\S#2#1#3}
\crefmultiformat{section}{\S\S#2#1#3}{, #2#1#3}{, #2#1#3}{, and~#2#1#3}

\onlineid{0}

\vgtccategory{Research}

\vgtcinsertpkg

\title{VizML: A Machine Learning Approach to\\Visualization Recommendation}

\author{Kevin Z. Hu, Michiel A. Bakker, Stephen Li, Tim Kraska, and C\'esar A. Hidalgo}
\authorfooter{
\item
    Kevin Z. Hu, Stephen Li, and C\'esar A. Hidalgo are with the Collective Learning Group at the MIT Media Lab. \{kzh, sli2014, hidalgo\}@mit.edu.
\item
    Michiel A. Bakker is with the Human Dynamics Group at the MIT Media Lab. bakker@mit.edu.
\item
    Tim Kraska is with the MIT Computer Science and Artificial Intelligence Laboratory. kraska@csail.mit.edu.
}

\abstract{
Data visualization should be accessible for all analysts with data, not just the few with technical expertise. Visualization recommender systems aim to lower the barrier to exploring basic visualizations by automatically generating results for analysts to search and select, rather than manually specify.  Here, we demonstrate a novel machine learning-based approach to visualization recommendation that learns visualization design choices from a large corpus of datasets and associated visualizations. First, we identify five key design choices made by analysts while creating visualizations, such as selecting a visualization type and choosing to encode a column along the X- or Y-axis. We train models to predict these design choices using one million dataset-visualization pairs collected from a popular online visualization platform. Neural networks predict these design choices with high accuracy compared to baseline models. We report and interpret feature importances from one of these baseline models. To evaluate the generalizability and uncertainty of our approach, we benchmark with a crowdsourced test set, and show that the performance of our model is comparable to human performance when predicting consensus visualization type, and exceeds that of other ML-based systems.
}
\keywords{Visualization Recommendation, Automated Visualization Design, Machine Learning}

\begin{document}

\firstsection{Introduction}

\maketitle

Knowledge workers across domains -- from business to journalism to scientific research -- increasingly use data visualization to generate insights, communicate, and make decisions~\cite{kandel-eda, narrative-visualization, rapid-adoption-data-driven-decision-making}. Yet, many visualization tools have steep learning curves due to a reliance on manual specification through code~\cite{ggplot2, d3} or clicks~\cite{polaris, spotfire}. As a result, data visualization is often inaccessible to the growing number of domain experts who lack the time or background to learn sophisticated tools.

While necessary to create bespoke visualizations, manual specification is overkill for many common use cases, such as preliminary data exploration and creating basic visualizations. To support these use cases in which speed and breadth of exploration are more important than customizability~\cite{tukey-eda}, systems can leverage the finding that \textit{the properties of a dataset influence how it can and should be visualized}. For example, prior research has shown that the accuracy with which visual channels (\textit{e.g.} position and color) encode data depends on the type~\cite{cleveland-mcgill-graphical-perception, Bertin:1983:SG:1095597, ware} and distribution~\cite{2018-task-data-effectiveness} of data values. 

Most recommender systems encode these visualization guidelines as collection of ``if-then" statements, or \textit{rules}~\cite{Hayes-Roth:1985:RS:4284.4286}, to automatically generate visualizations for analysts to search and select, rather than manually specify~\cite{toward-visualization-recommendation}. For example, APT~\cite{automating-the-design}, BOZ~\cite{boz}, and SAGE~\cite{sage} generate and rank visualizations using rules informed by perceptual principles. Recent systems such as Voyager~\cite{2016-voyager, 2017-voyager2}, Show Me~\cite{show-me}, and DIVE~\cite{dive-2018} extend these approaches with support for column selection. While effective for certain use cases~\cite{2016-voyager}, these \textit{rule-based} approaches face limitations such as the cold-start problem, costly rule creation, and the combinatorial explosion of results~\cite{recommender-systems-the-book}.

In contrast, \textit{machine learning (ML)-based} systems directly learn the relationship between data and visualizations by training models on analyst interaction. While recent systems like DeepEye~\cite{deepeye}, Data2Vis~\cite{data2vis}, and Draco-Learn~\cite{draco} are exciting, they do not learn to make visualization design choices as an analyst would, which impacts interpretability and ease of integration into existing systems. Furthermore, because these systems are trained with crowdsourced annotations on rule-generated visualizations in controlled settings, they are limited by the quantity and quality of data.

\pagebreak

We introduce \textbf{VizML}, a ML-based approach to visualization recommendation using a large corpus of datasets and associated visualizations. To begin, we describe visualization as a process of making design choices that maximize effectiveness, which depends on dataset, task, and context. Then, we formulate visualization recommendation as a problem of developing models that learn make design choices.

We train and test these models using one million unique dataset-visualization pairs from the Plotly Community Feed~\cite{plotly-community-feed}. We describe our process of collecting and cleaning this corpus, extracting features from each dataset, and extracting seven key design choices from corresponding visualizations. Our learning tasks are to optimize models that use features to predict these choices.

Neural networks trained on 60\% of the corpus achieve $\sim70-95\%$ accuracy at predicting design choices in a separate 20\% test set. This performance exceeds that of four simpler baseline models, which themselves out-perform random chance. We report feature importances from one of these baseline models, interpret the contribution of features to a given task, and relate them to existing research. 

We evaluate the generalizability and uncertainty of our model by benchmarking against a crowdsourced test set. We construct this test set by randomly selecting datasets from Plotly, visualizing each as a bar, line, and scatter plot, and measuring the consensus of Mechanical Turk workers. Using a scoring metric that adjusts for the degree of consensus, we find that VizML performs comparably to Plotly users and Mechanical Turkers, and outperforms other ML-based models.

To conclude, we discuss interpretations, limitations, and extensions of our initial machine learning approach to visualization recommendation. We also suggest directions for future research, such as integrating separate recommender models into an end-to-end system, developing public benchmarking corpuses, and employing unsupervised models.

\begin{mdframed}[backgroundcolor=black!10]
    Key contributions:
    \begin{enumerate}[noitemsep, leftmargin=*, topsep=3pt]
        \item \textbf{Problem formulation} (\cref{sec:problem}): learning design choices from a corpus of data-visualization pairs
        \item \textbf{Data processing pipeline} (\cref{sec:data,sec:methods}): collecting and cleaning corpus, then extracting features and design choices
        \item \textbf{Predicting design choices} (\cref{sec:results}): evaluating neural network performance at predicting design choices
        \item \textbf{Feature importances} (\cref{sec:features}): reporting and interpreting the contribution of each feature to the prediction tasks
        \item \textbf{Crowdsourced benchmark} (\cref{sec:ground-truth}): evaluating human and ML models at predicting crowdsourced visualization type 
    \end{enumerate}
\end{mdframed}
\pagebreak
\section{Problem Formulation}\label{sec:problem}
Data visualization communicates information by representing data with visual elements. These representations are specified using \textit{encodings} that map from data to the \textit{retinal properties} (\textit{e.g.} position, length, or color) of \textit{graphical marks} (\textit{e.g.} points, lines, or rectangles)~\cite{Bertin:1983:SG:1095597, Card:1999:RIV:300679}.

Concretely, consider a dataset that describes 406 automobiles (rows) with eight attributes (columns) such as miles per gallon (\texttt{MPG}), horsepower (\texttt{Hp}), and weight in pounds (\texttt{Wgt}) ~\cite{cars}. To create a scatterplot showing the relationship between \texttt{MPG} and \texttt{Hp}, an analyst encodes each pair of data points with the position of a circle on a 2D plane, while also specifying many other properties like size and color:

{\setlength\intextsep{2pt}
    \begin{figure}[ht]
     \centerline{\includegraphics[keepaspectratio]{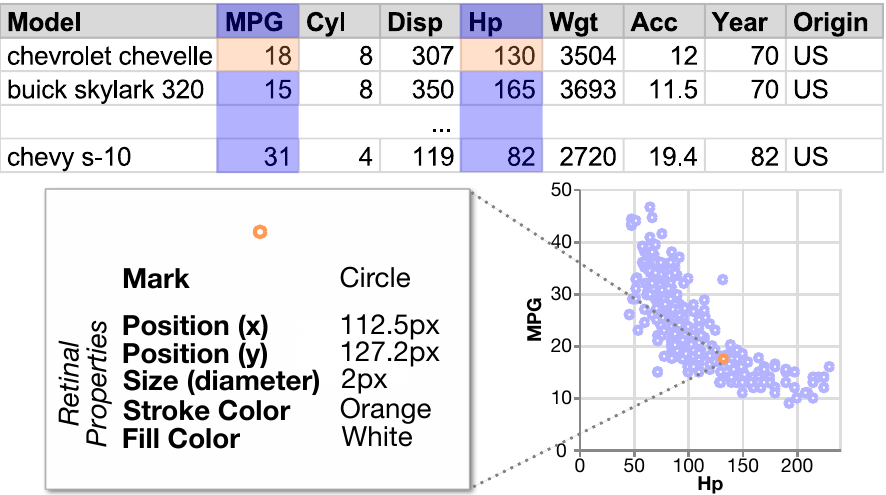}}
    \end{figure}
}

To create bespoke visualizations, analysts might need to exhaustively specify encodings in detail using expressive tools. But a scatterplot is specified with the Vega-lite~\cite{vegalite} grammar by selecting a mark type and fields to be encoded along the x- and y-axes, and in Tableau~\cite{polaris} by placing the two columns onto the respective column and row shelves.

{\setlength\intextsep{2pt}
    \begin{figure}[ht]
     \centerline{\includegraphics[]{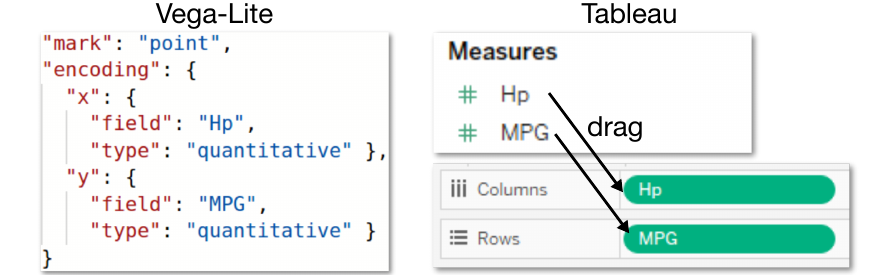}}
    \end{figure}
}

That is, to create basic visualizations in many grammars or tools, an analyst specifies higher-level \textit{design choices}, which we define as statements that compactly and uniquely specify a bundle of lower-level encodings. Equivalently, each grammar or tool affords a design space of visualizations, which a user constrains by making choices.

\subsection{Visualization as Making Design Choices}
We formulate basic visualization of a dataset $d$ as a set of interrelated design choices $C=\{c\}$, each of which is selected from a possibility space $c \sim \mathbb{C}$. However, not all design choices result in valid visualizations -- some choices are incompatible with each other. For instance, encoding a categorical column with the Y position of a line mark is invalid. Therefore, the set of choices that result in valid visualizations is a subset of the space of all possible choices $\mathbb{C}_1 \times \mathbb{C}_2 \times \ldots \times \mathbb{C}_{|C|}$.

\begin{figure}[ht]
 \includegraphics[width=\columnwidth]{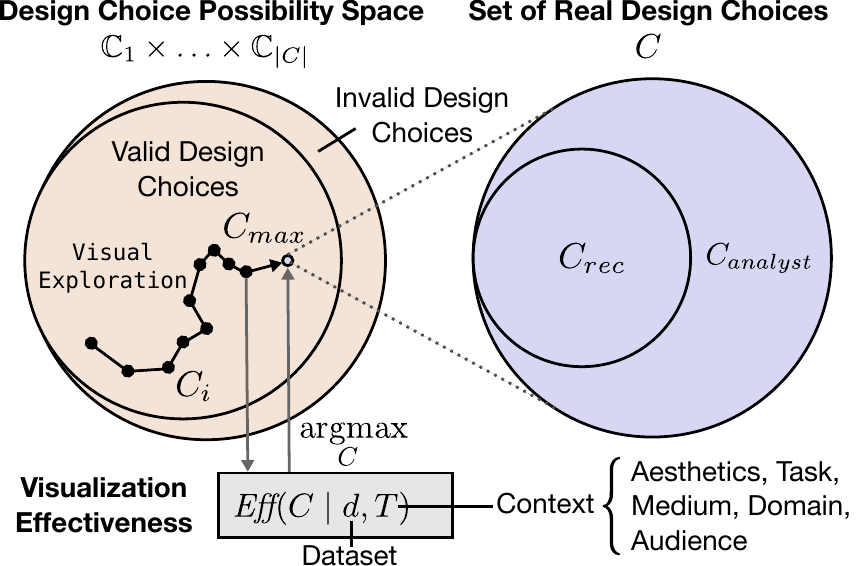}
 \caption{Creating visualizations is a process of making design choices, which can be recommended by a system or specified by an analyst.}
 \label{fig:design-choices}
\end{figure}

The effectiveness of a visualization can be defined by informational measures such as efficiency, accuracy, and memorability~\cite{measuring-effective-data-visualization, memorability}, or emotive measures like engagement~\cite{engaging-with-big-data-visualizations, data-visualization-effectiveness-profile}. Prior research also shows that effectiveness is informed by low-level perceptual principles~\cite{cleveland-mcgill-graphical-perception, somewhere-over-the-rainbow, sizing-the-horizon, graphical-perception-maps} and dataset properties~\cite{evaluating-visualization-techniques-and-tools, 2018-task-data-effectiveness}, in addition to contextual factors such as task~\cite{task-based-effectiveness, 2018-task-data-effectiveness, low-level-components-of-analytic-activity}, aesthetics~\cite{aesthetic-usability}, domain~\cite{improving-visualization-by-capturing-domain-knowledge}, audience~\cite{color-in-visualization}, and medium~\cite{data-in-vr, text-2d-3d-interfaces}. In other words, an analyst makes design choices $C_{max}$ that maximize visualization effectiveness $\textit{Eff}$ given a dataset $d$ and contextual factors $T$:

\begin{equation}
    C_{max} = \argmax_C \textit{Eff}(C \mid d, T)
    \label{eq:max-eff}
\end{equation}
But making design choices can be expensive. A goal of visualization recommendation is to reduce the cost of creating visualizations by automatically suggesting a subset of design choices $C_{rec} \subseteq C$.

\subsection{Modeling Design Choice Recommendation}

Consider a single design choice $c \in C$. Let $C' = C \setminus{\{c\}}$ denote the set of all other design choices excluding $c$. Given $C'$, a dataset $d$, and context $T$, there is an ideal design choice recommendation function $F_c$ that outputs the design choice $c_{max} \in C_{max}$ from Eqn.~\ref{eq:max-eff} that maximizes visualization effectiveness:

\begin{equation}
F_c(d \mid C', T) = c_{max}
\end{equation}

Our goal is to approximate $F_c$ with a function $\textit{G}_c \approx F_c$. Assume now a corpus of datasets $D=\{d\}$ and corresponding visualizations $V=\{V_d\}$, each of which can be described by design choices $C_d=\{c_d\}$. Machine learning-based recommender systems consider $G_c$ as a model with a set of parameters $\Theta_c$ that can be trained on this corpus by a learning algorithm that maximizes an objective function $\textit{Obj}$:

\begin{equation}
    \Theta_{fit} = \argmax_{\Theta_c} \sum_{d \in D} \textit{Obj}(c_d, G_c(d \mid \Theta_c, C', T))
    \label{eq:max-acc}
\end{equation}

Without loss of generality, say the objective function maximizes the likelihood of observing the training output $\{C_d\}$. Even if an analyst makes sub-optimal design choices, collectively optimizing the likelihood of all observed design choices can still be optimal~\cite{learning-from-noisy-labels}. This is precisely the case with our observed design choices $c_d = F_c(d \mid C', T) + \text{noise} + \text{bias}$. Therefore, given an unseen dataset $d^{*}$, maximizing this objective function can plausibly lead to a recommendation that maximizes effectiveness of a visualization. 

\begin{equation}
    G_c( d^{*} \mid \Theta_{fit}, C', T) \approx F_c(d^{*} \mid C', T) = c_{max}
\end{equation}

In this paper, our model $G_c$ is a neural network and $\Theta_c$ are connection weights. We simplify the recommendation problem by optimizing each $G_c$ independently, and without contextual factors: $G_c(d \mid \Theta )=G_c( d \mid \Theta, C', T)$. We note that independent recommendations may not be compatible, nor do they necessarily maximize overall effectiveness. Generating a complete visualization output will require modeling dependencies between $G_c$ for each $c$, which we discuss in~\cref{sec:future-work}.

\begin{figure}[ht]
    \includegraphics[width=\columnwidth]{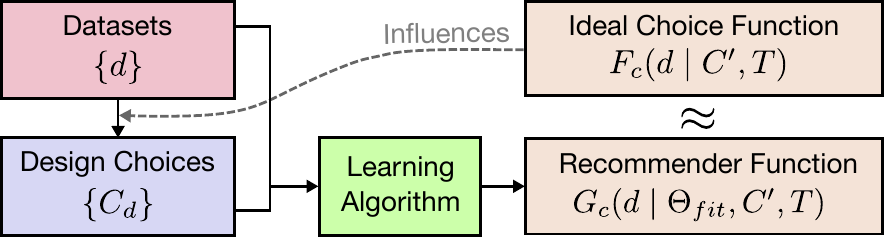}
    \caption{Basic setup of learning models to recommend design choices with a corpus of datasets and corresponding design choices.}
    \label{fig:modelling}
\end{figure}

\begin{table*}[t!]
    \resizebox{\textwidth}{!}{
    \begin{tabular}{@{}llllllll@{}}
    \toprule
    \textbf{System} & \textbf{Source} & \textbf{$\mathbf{N_{data}}$} & \textbf{Generation} & \textbf{Learning Task} & \textbf{Training Data} & \textbf{Features} & \textbf{Model} \\ \midrule
    \multirow{2}{*}{VizML} & \multirow{2}{*}{\begin{tabular}[c]{@{}l@{}}Public\\ (Plotly)\end{tabular}} & \multirow{2}{*}{$10^6$} & \multirow{2}{*}{Human} & \multirow{2}{*}{\begin{tabular}[c]{@{}l@{}}Design Choice\\ Recommendation\end{tabular}} & \multirow{2}{*}{\begin{tabular}[c]{@{}l@{}}Dataset-\\ Visualization Pairs\end{tabular}} & \multirow{2}{*}{\begin{tabular}[c]{@{}l@{}}Single + Pairwise +\\Aggregated\end{tabular}} & \multirow{2}{*}{Neural Network} \\
     &  &  &  &  &  &  & \\ \midrule
    \multirow{2}{*}{DeepEye} & \multirow{2}{*}{Crowd} & \multirow{2}{*}{\begin{tabular}[c]{@{}l@{}}1) 33.4K\\2) 285K\end{tabular}} & \multirow{2}{*}{\begin{tabular}[c]{@{}l@{}}Rules $\rightarrow$\\ Annotation\end{tabular}} & \multirow{2}{*}{\begin{tabular}[c]{@{}l@{}}1) Good-Bad Classif.\\ 2) Ranking\end{tabular}} & \multirow{2}{*}{\begin{tabular}[c]{@{}l@{}}1) Good-Bad Labels\\ 2) Pairwise Comparisons\end{tabular}} & \multirow{2}{*}{Column Pair} & \multirow{2}{*}{\begin{tabular}[c]{@{}l@{}}1) Decision Tree\\ 2) RankNet\end{tabular}} \\
     &  &  &  &  &  &  & \\ \midrule
    \multirow{2}{*}{Data2Vis} & \multirow{2}{*}{\begin{tabular}[c]{@{}l@{}}Tool\\ (Voyager)\end{tabular}} & \multirow{2}{*}{4,300} & \multirow{2}{*}{\begin{tabular}[c]{@{}l@{}}Rules $\rightarrow$\\ Validation\end{tabular}} & \multirow{2}{*}{\begin{tabular}[c]{@{}l@{}}End-to-End\\ Viz. Generation\end{tabular}} & \multirow{2}{*}{\begin{tabular}[c]{@{}l@{}}Dataset Subset-\\ Visualization Pairs\end{tabular}} & \multirow{2}{*}{Raw} & \multirow{2}{*}{Seq2Seq NN} \\
     &  &  &  &  &  &  & \\\midrule
    \multirow{2}{*}{Draco-Learn} & \multirow{2}{*}{\begin{tabular}[c]{@{}l@{}}Crowd\end{tabular}} & \multirow{2}{*}{\begin{tabular}[c]{@{}l@{}}1,100 + \\10\end{tabular}} & \multirow{2}{*}{\begin{tabular}[c]{@{}l@{}}Rules $\rightarrow$\\ Annotation\end{tabular}} & \multirow{2}{*}{\begin{tabular}[c]{@{}l@{}}Soft Constraint\\Weights\end{tabular}} & \multirow{2}{*}{\begin{tabular}[c]{@{}l@{}}Pairwise Comparisons\end{tabular}} & \multirow{2}{*}{\begin{tabular}[c]{@{}l@{}}Soft Constraint\\Violation Counts\end{tabular}} & \multirow{2}{*}{RankSVM} \\
     &  &  &  &  &  &  & \\ \bottomrule
    \end{tabular}
}

\caption{Comparison of machine learning-based visualization recommendation systems. The major differences are that of \textbf{Learning Task} definition, and the quantity ($\mathbf{N_{data}}$) and quality (\textbf{Generation} and \textbf{Training Data}) of training data.}
\label{tab:ml-based-system-comparison}

\end{table*}

\section{Related Work}\label{sec:related}
We relate and compare our work to existing \textit{Rule-based Visualization Recommender Systems}, \textit{ML-based Visualization Recommender Systems}, and prior \textit{Descriptions of Public Data and Visualizations}.

\subsection{Rule-based Visualization Recommender Systems}

Visualization recommender systems either suggest data queries (selecting \textit{what} data to visualize) or visual encodings (\textit{how} to visualize selected data)~\cite{2016-compassql}. Data query recommenders vary widely in their approaches~\cite{rank-by-feature, wilkinson-scagnostics}, with recent systems optimizing statistical ``utility" functions~\cite{seedb, muve}. Though specifying data queries is crucial to visualization, it is distinct from design choice recommendation.

Most visual encoding recommenders implement guidelines informed the seminal work of Bertin~\cite{Bertin:1983:SG:1095597}, Cleveland and McGill~\cite{cleveland-mcgill-graphical-perception}, and others. This approach is exemplified by Mackinlay's \textbf{APT}~\cite{automating-the-design} -- the \textit{ur-}recommender system -- which enumerates, filters, and scores visualizations using \textit{expressiveness} and perceptual \textit{effectiveness} criteria. The closely related \textbf{SAGE}~\cite{sage}, \textbf{BOZ}~\cite{boz}, and \textbf{Show Me}~\cite{show-me} support more data, encoding, and task types. Recently, hybrid systems such as \textbf{Voyager}~\cite{2016-voyager, 2017-voyager2, 2016-compassql}, \textbf{Explore in Google Sheets}~\cite{google-sheets-explore, google-sheets-explore-patent}, \textbf{VizDeck}~\cite{perry2013vizdeck}, and \textbf{DIVE}~\cite{dive-2018} combine visual encoding rules with the recommendation of visualizations that include non-selected columns.

Though effective for many use cases, these systems suffer from four major limitations. First, visualization is a complex process that may require encoding non-linear relationships that are difficult to capture with simple rules. Second, even crafting simple rule sets is a costly process that relies on expert judgment. Third, like rule-based systems in other domains, these systems face the \textit{cold-start problem} of presenting non-trivial results for datasets or users about which they have not yet gathered sufficient information~\cite{recommender-systems-the-book}. Lastly, as the dimension of input data increases, the combinatorial nature of rules result in an explosion of possible recommendations.

\subsection{ML-based Visualization Recommender Systems}

The guidelines encoded by rule-based systems often derive from experimental findings and expert experience. Therefore, an indirect manner, heuristics distill best practices learned from another analyst's experience while creating or consuming visualizations. Instead of aggregating best practices learned from data, and representing them in a system with rules, ML-based systems propose to train models that learn directly from data, and can be embedded into systems \textit{as-is}.

\textbf{DeepEye}~\cite{deepeye} combines rule-based visualization generation with models trained to 1) classify a visualization as ``good" or ``bad" and 2) rank lists of visualizations. The DeepEye corpus consists of 33,412 bivariate visualizations of columns drawn from 42 public datasets. 100 students annotated these visualizations as good/bad, and compared 285,236 pairs. These annotations, combined with 14 features for each column pair, train a decision tree for the classification task and a ranking neural network~\cite{ranknet} for the ``learning to rank" task.

\textbf{Data2Vis}~\cite{data2vis} uses a neural machine translation approach to create a sequence-to-sequence model that maps JSON-encoded datasets to Vega-lite visualization specifications. This model is trained using 4,300 automatically generated Vega-Lite examples, consisting of 1-3 variables, generated from 11 distinct datasets. The model is qualitatively validated by examining the visualizations generated from 24 common datasets.

\textbf{Draco-Learn}~\cite{draco} learns trade-offs between constraints in Draco, a formal model that represents 1) visualizations as logical facts and 2) design guidelines as hard and soft constraints.  Draco-Learn uses a ranking support vector machine trained on ranked pairs of visualizations harvested from graphical perception studies~\cite{2018-task-data-effectiveness, task-based-effectiveness}. Draco can recommend visualizations that satisfy these constraints by solving a combinatorial optimization problem.

VizML differs from these systems in three major respects, as shown in Table~\ref{tab:ml-based-system-comparison}. In terms of the \textit{learning task}, DeepEye learns to classify and rank visualizations, Data2Vis learns an end-to-end generation model, and Draco-Learn learns soft constraints weights. By learning to predict design choices, VizML models are easier to quantitatively validate, provide interpretable measures of feature importance, and can be more easily integrated into visualization systems. 

In terms of \textit{data quantity}, the VizML training corpus is orders of magnitude larger than that of DeepEye and Data2Vis. The size of our corpus permits the use of 1) large feature sets that capture many aspects of a dataset and 2) high-capacity models like deep neural networks that can be evaluated against a large test set.

The third major difference is one of \textit{data quality}. The datasets used to train VizML models are extremely diverse in shape, structure, and other properties, in contrast to the few datasets used to train the three existing systems. Furthermore, the visualizations used by other ML-based recommender systems are still generated by rule-based systems, and evaluated in controlled settings. The corpus used by VizML is the result of real visual analysis by analysts on their own datasets.

However, VizML faces two major limitations. First, these three ML-based systems recommend both data queries and visual encodings, while VizML only recommends the latter. Second, in this paper, we do not create an application that employs our visualization model. Design considerations for user-facing systems that productively and properly employ ML-based visualization recommendation are important, but beyond the scope of this paper.

\subsection{Descriptions of Public Data and Visualizations}

\textbf{Beagle}~\cite{battle-beagle} is an automated system for scraping over $41,000$ visualizations across five tools from the web. Beagle shows that a few visualization types represent a large portion of visualizations, and shows difference in visualization type usage between tools. However, Beagle does not collect the data used to generate these visualizations.

A \textbf{2013 study of ManyEyes and Tableau Public}~\cite{public-data-and-visualizations} analyzes hundreds of thousands of datasets and visualizations from two popular tools~\cite{manyeyes, polaris}. The authors report usage patterns, distribution of dataset properties, and characteristics of visualizations. This study also relates dataset properties with visualization types, similar to predicting visualization type using dimension-based features in our approach.
\section{Data}\label{sec:data}
We describe our process for collecting and cleaning a corpus of 2.3 million dataset-visualization pairs, describing each dataset and column with features, and extracting design choices from each visualization. These are steps \textbf{1}, \textbf{2}, and \textbf{3} of the workflow shown in Fig.~\ref{fig:processing-flow}.
\subsection{Collection and Cleaning}

\textbf{Plotly}~\cite{plotly} is a software company that creates tools and software libraries for data visualization and analysis. For example, Plotly Chart Studio~\cite{plotly-chart-studio} is a web application that lets users upload datasets and manually create interactive D3.js and WebGL visualizations of over 20 visualization types. Users familiar with Python can use the Plotly Python library~\cite{plotly-python-library} to create those same visualizations with code.

Visualizations in Plotly are specified with a declarative schema. In this schema, each visualization is specified with two data structures. The first is a list of \textit{traces} that specify how a collection of data is visualized. The second is a dictionary that specifies aesthetic aspects of a visualization untied from the data, such as axis labels and annotations. For example, the scatterplot from~\cref{sec:problem} is specified with a single ``scatter" trace with \texttt{Hp} as the \texttt{x} parameter and \texttt{MPG} as the \texttt{y} parameter:

{\setlength\intextsep{5pt}
\begin{figure}[htbp]
    \includegraphics[width=\columnwidth]{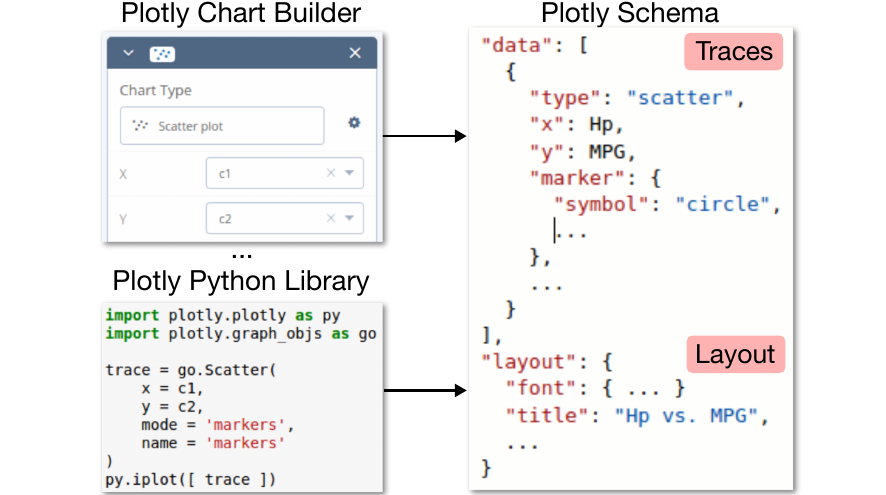}
\end{figure}
}

The Plotly schema is similar to that of MATLAB and of the matplotlib Python library. The popular Vega~\cite{2014-reactive-vega} and Vega-lite~\cite{vegalite} schemas are more opinionated, which ``allows for complicated chart display with a concise JSON description, but leaves less control to the user"~\cite{plotly-open-source}. Despite these differences, it is straightforward to convert Plotly schemas into other schemas, and vice versa.

\begin{figure}[b!]
 \includegraphics[width=\columnwidth]{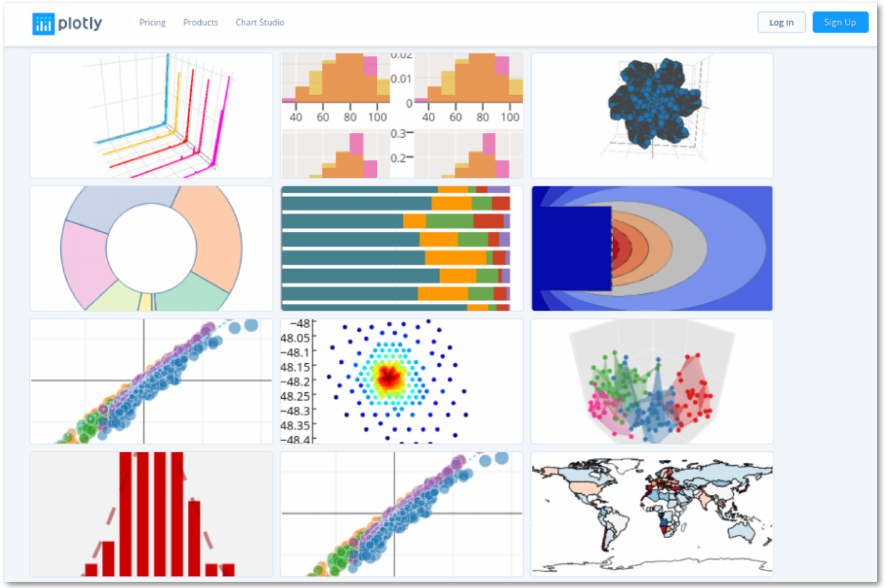}
 \caption{Screenshot of the Plotly Community Feed~\cite{plotly-community-feed}.}
 \label{fig:plotly-community-feed}
\end{figure}

Plotly also supports sharing and collaboration. Starting in 2015, users could publish charts to the Plotly Community Feed~\cite{plotly-community-feed}, which provides an interface for searching, sorting, and filtering millions of visualizations, as shown in Fig.~\ref{fig:plotly-community-feed}. The underlying \texttt{/plots} endpoint from the Plotly REST API~\cite{plotly-rest-api} associates each visualization with three objects: \texttt{data} contains the source data, \texttt{specification} contains the traces, and \texttt{layout} defines display configuration. 

\subsection{Data Description}\label{subsec:data-description}
Using the Plotly API, we collected approximately 2.5 years of public visualizations from the feed, starting from 2015-07-17 and ending at 2018-01-06. We gathered 2,359,175 visualizations in total, 2,102,121 of which contained all three configuration objects, and 1,989,068 of which were parsed without error. To avoid confusion between user-uploaded datasets and our dataset of datasets, we refer to this collection of dataset-visualization pairs as the \textit{Plotly corpus}.

\begin{figure}[hb]
 \includegraphics[width=\columnwidth]{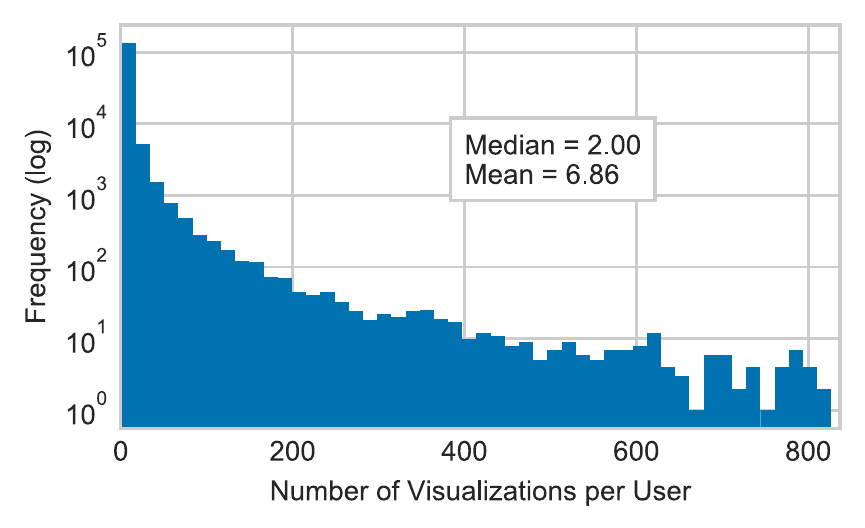}
 \vspace{-0.7cm}
    \caption{Distribution of plots per user, visualized on a log-log scale.}
    \label{fig:plots-per-user}
\end{figure}

The Plotly corpus contains visualizations created by $143,007$ unique users, who vary widely in their usage. The distribution of visualizations per user is shown in Fig.~\ref{fig:plots-per-user}. Excluding the top $0.1\%$ of users with the most visualizations, many of whom are bots that programmatically generate visualizations, users created a mean of $6.86$ and a median of $2$ visualizations each.

\begin{figure}[hb]
    \begin{subfigure}{\columnwidth}
        \includegraphics[width=\columnwidth]{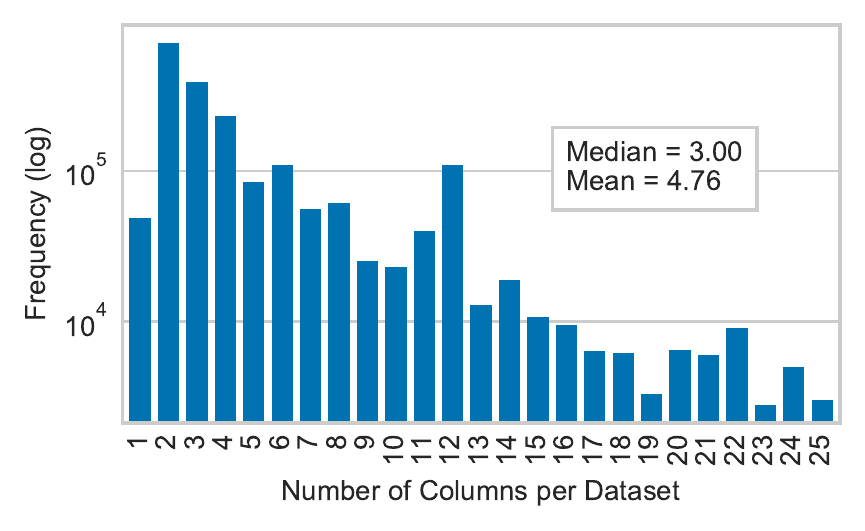}
        \vspace{-0.7cm}
        \caption{Distribution of columns per dataset, after removing the $5.03\%$ of datasets with more than 25 columns, visualized on a log-linear scale.}
        \label{fig:columns-per-dataset}
    \end{subfigure}

    \begin{subfigure}{\columnwidth}
        \includegraphics[width=\columnwidth]{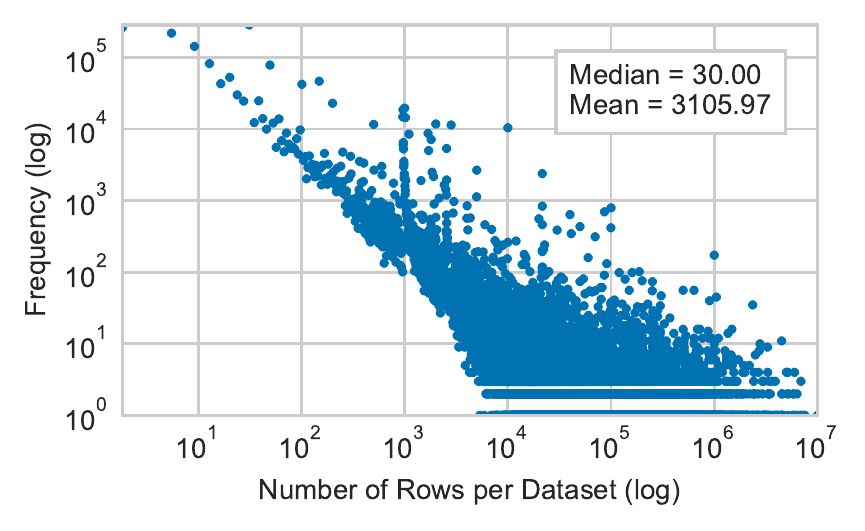}
        \vspace{-0.7cm}
        \caption{Distribution of rows per dataset, visualized on a log-log scale.}
        \label{fig:rows-per-dataset}
    \end{subfigure} 

\caption{Distribution of dataset dimensions in the Plotly corpus.}
\end{figure}

Datasets also vary widely in number of columns and rows. Though some datasets contain upwards of $100$ columns, $94.97\%$ contain less than or equal to $25$ columns. Excluding datasets with more than $25$ columns, the average dataset has $4.75$ columns, and the median dataset has $3$ columns. The distribution of columns per visualization is shown in Fig.~\ref{fig:columns-per-dataset}. The distribution of rows per dataset is shown in Fig.~\ref{fig:rows-per-dataset}, and has a mean of $3105.97$, median of $30$, and maximum of $10 \times 10^6$. These heavy-tailed distributions are consistent with those of IBM ManyEyes and Tableau Public as reported by~\cite{public-data-and-visualizations}.

Though Plotly lets users generate visualizations using multiple datasets, $98.32\%$ of visualizations used only one source dataset. Therefore, we are only concerned with visualizations using a single dataset. Furthermore, over $90\%$ of visualizations used all columns in the source dataset, so we are not able to address data query selection. Lastly, out of $13,321,598$ traces, only $0.16\%$ of have transformations or aggregations. Given this extreme class imbalance, we are not able to address column transformation or aggregation as learning tasks.

\subsection{Feature Extraction}\label{sec:feature-extraction}
We describe each column with the \textbf{81 single-column features} shown in Table~\ref{tab:features-list_single-column} in the Appendix. These features fall into four categories. The \textbf{Dimensions (D)} feature is the number of rows in a column. \textbf{Types (T)} features capture whether a column is categorical, temporal, or quantitative. \textbf{Values (V)} features describe the statistical and structural properties of the values within a column. \textbf{Names (N)} features describe the column name.

We distinguish between these feature categories for three reasons. First, these categories let us organize how we create and interpret features. Second, we can observe the contribution of different types of features. Third, some categories of features may be less generalizable than others. We order these categories (\textbf{D} $\rightarrow$ \textbf{T} $\rightarrow$ \textbf{V} $\rightarrow$ \textbf{N}) by how biased we expect those features to be towards the Plotly corpus.

Nested within these categories are more groupings of features. For instance, within the \textbf{Values} category, the \textit{Sequence} group includes measures of sortedness, while the features within the \textit{Unique} group describes the uniqueness of values in a column.

We describe each pair of columns with \textbf{30 pairwise-column features}. These features fall into two categories: \textbf{Values} and \textbf{Names}, some of which are shown in Table~\ref{tab:features-list_pairwise-column}. Note that many pairwise-column features, depend on the individual column types determined through single-column feature extraction. For instance, the Pearson correlation coefficient requires two numeric columns, and the ``number of shared values" feature requires two categorical columns.

We create \textbf{841 dataset-level features} by aggregating these single- and pairwise-column features using the \textbf{16 aggregation functions} shown in Table~\ref{tab:aggregations}. These aggregation functions convert single-column features (across all columns) and pairwise-column features (across all pairs of columns) into scalar values. For example, given a dataset, we can count the number of columns, describe the percent of columns that are categorical, and compute the mean correlation between all pairs of quantitative columns. Two other approaches to incorporating single-column features are to train separate models per number of columns, or to include column features with padding. Neither approach yielded a significant improvement over the results in \cref{sec:results}.

\begin{figure}[t]
 \includegraphics[width=\columnwidth]{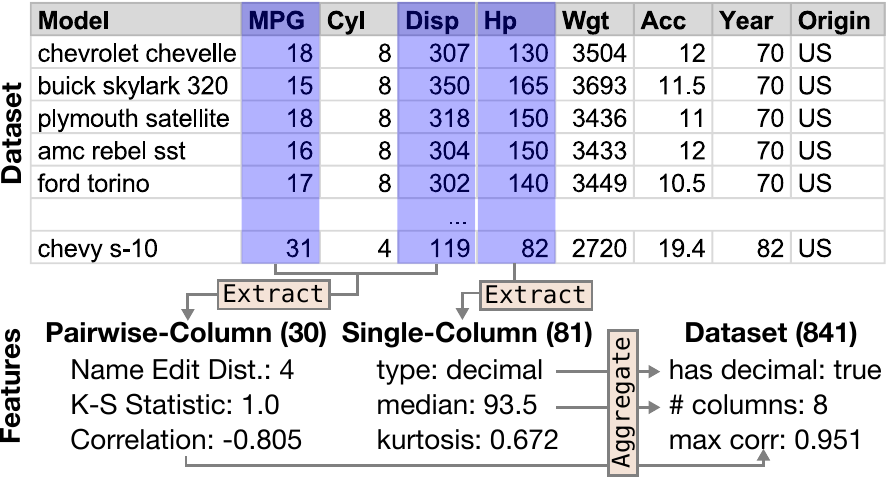}
    \caption{Extracting features from the Automobile MPG dataset~\cite{cars}.}
    \label{fig:feature-extraction}
\end{figure}

\subsection{Design Choice Extraction}\label{sec:chart-outcomes}

\begin{figure}[t]
\includegraphics[width=\columnwidth]{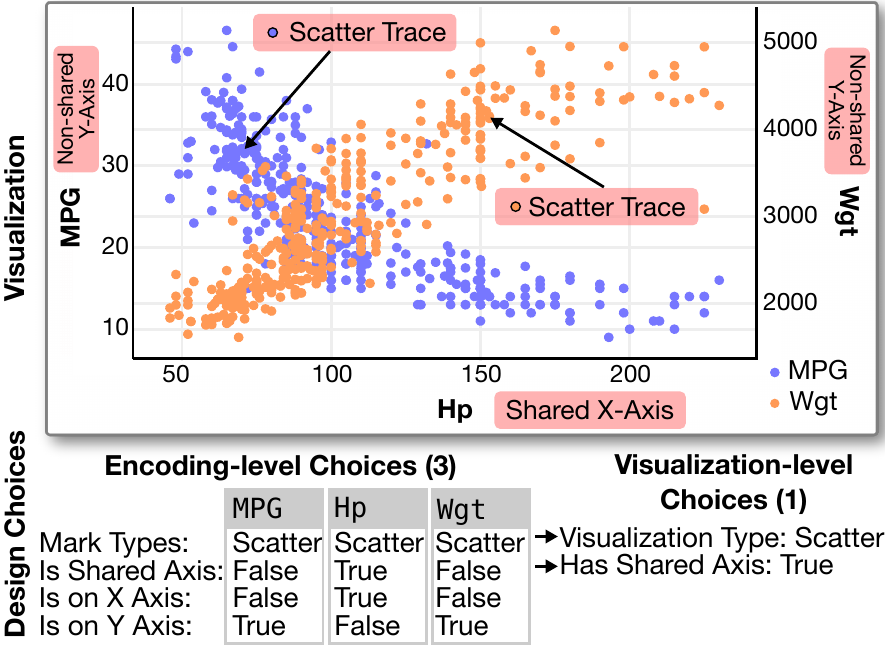}
\caption{Extracting design choices from a dual-axis scatterplot.}
\label{fig:outcome-extraction}
\end{figure}

Each visualization in Plotly consists of traces that associate collections of data with visual elements. Therefore, we extract an analyst's design choices by parsing these traces. Examples of \textbf{encoding-level design choices} include \textit{mark type}, such as scatter, line, bar; and \textit{X or Y column encoding}, which specifies which column is represented on which axis; and whether or not an X or Y column is the single column represented along that axis. For example, the visualization in Fig.~\ref{fig:outcome-extraction} consists of two scatter traces, both of which have the same column encoded on the X axis (\texttt{Hp}), and two distinct columns encoded on the Y axis (\texttt{MPG} and \texttt{Wgt}).

By aggregating these encoding-level design choices, we can characterize \textbf{visualization-level design choices} of a chart. Within our corpus, over $90\%$ of the visualizations consist of homogeneous mark types. Therefore, we use \textit{visualization type} to describe the type shared among all traces, and also determined whether the visualization \textit{has a shared axis}. The example in Fig.~\ref{fig:outcome-extraction} has a scatter  visualization type and a single shared axis (X).

\begin{SCfigure*}[][t!]
    \includegraphics[width=1.5\columnwidth]{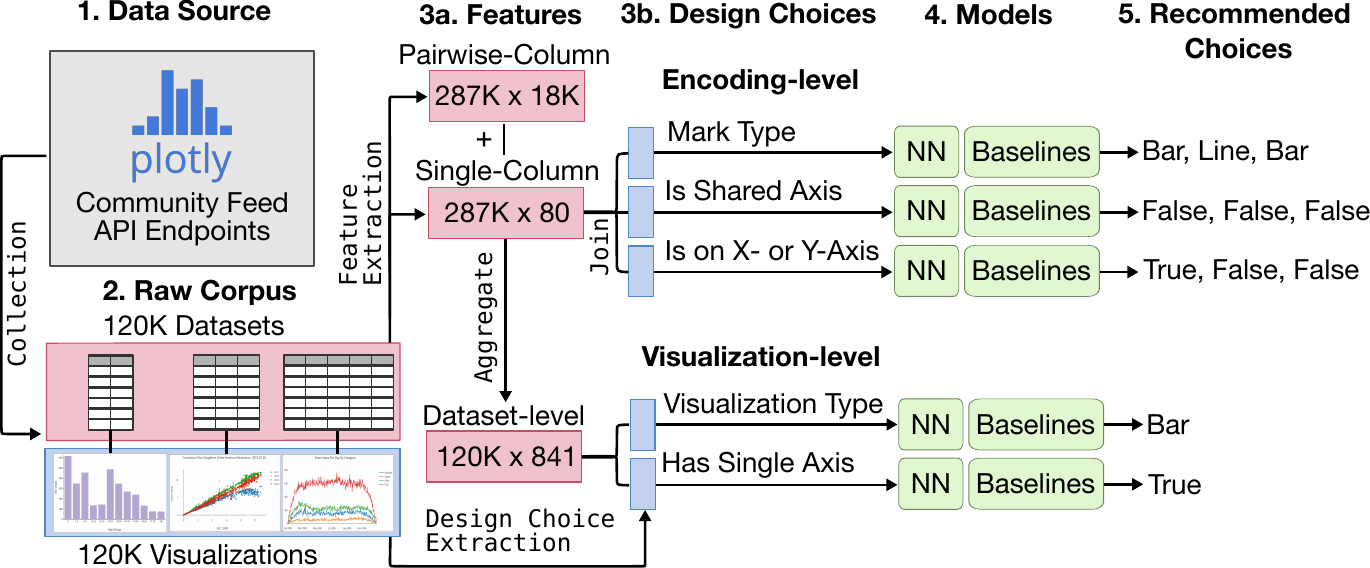} 
    \caption{Diagram of data processing and analysis flow, starting from (\textbf{1}) the original Plotly API endpoints, proceeding to (\textbf{2}) the deduplicated dataset-visualization pairs, (\textbf{3a}) the features describing each individual column, pair of columns, and dataset, (\textbf{3b}) design choices extracted from visualizations extraction process, (\textbf{4}) the task-specific models trained on these features, (\textbf{5}) and predicted choices. }
    \label{fig:processing-flow}
\end{SCfigure*}
\section{Methods}\label{sec:methods}

We describe our feature processing pipeline, the machine learning models we use, how we train those models, and how we evaluate performance. These are steps \textbf{4} and \textbf{5} of the workflow in Fig.~\ref{fig:processing-flow}.

\subsection{Feature Processing}
We converted raw features into a form suitable for modeling with a five-stage pipeline. First, we apply one-hot encoding to categorical features. Second, we set numeric values above the 99th percentile or below the 1st percentile to those respective cut-offs. Third, we imputed missing categorical values using the mode of non-missing values, and missing numeric values with the mean of non-missing values. Fourth, we removed the mean of numeric fields and scaled to unit variance.

Lastly, we randomly removed datasets that were exact deduplicates of each other, resulting in unique $1,066,443$ datasets and $2,884,437$ columns. However, many datasets were slight modifications of each other, uploaded by the same user. Therefore, we removed all but one randomly selected dataset per user, which also removed bias towards more prolific Plotly users. This aggressive deduplication resulted in a final corpus of \textbf{119,815 datasets} and \textbf{287,416 columns}. Results from only exact deduplication result in significantly higher within-corpus test accuracies, while a soft threshold-based deduplication results in similar test accuracies.  

\subsection{Prediction Tasks}
Our task is to train models that use the features described in~\cref{sec:feature-extraction} to predict the design choices described in~\cref{sec:chart-outcomes}. \textbf{Two visualization-level prediction tasks} use dataset-level features to predict visualization-level design choices:

\begin{mdframed}[backgroundcolor=black!10, nobreak=true]
\begin{enumerate}[leftmargin=*]
\setlength\itemsep{0.2em}
  \item \textbf{Visualization Type [VT]}: \textit{2-, 3-, and 6-class}\\
  Given that all traces are the same type, what type is it?
  
  \vspace*{-0.5\baselineskip}
  \begin{tabular}{@{}llllll@{}}
    \textit{Scatter} & \textit{Line} & \textit{Bar} & \textit{Box} &  \textit{Histogram} & \textit{Pie} \\
    44829 & 26209 & 16002 & 4981 & 4091 & 3144 
  \end{tabular}
  
  \item \textbf{Has Shared Axis [HSA]}: \textit{2-class}\\
  Do the traces in the chart all share one axis (either X or Y)? 

  \vspace*{-0.5\baselineskip}
  \begin{tabular}{@{}ll@{}}
    \textit{False} & \textit{True} \\
    95723 & 24092
  \end{tabular}
\end{enumerate}
\end{mdframed}

The \textbf{three encoding-level prediction tasks} use features about individual columns to predict how they are visually encoded. That is, these prediction tasks consider each column independently, instead of alongside other columns in the same dataset. This \textit{bag-of-columns} approach accounts for the effect of column order.

\begin{mdframed}[backgroundcolor=black!10, nobreak=true]
 \begin{enumerate}[leftmargin=*]
  \setlength\itemsep{0.2em}
  \item \textbf{Mark Type [MT]}: \textit{2-, 3-, and 6-class}\\
  What mark type is used to represent this column?
  
  \vspace*{-0.5\baselineskip}
  \begin{tabular}{@{}llllll@{}}
    \textit{Scatter} & \textit{Line} & \textit{Bar} & \textit{Box} &  \textit{Histogram} & \textit{Heatmap} \\
    68931 & 64726 & 30023 & 13125 & 5163 & 1032
  \end{tabular}
  
  \item \textbf{Is Shared X-axis or Y-axis [ISA]}: \textit{2-class}\\
  Is this column the only column on encoded on its axis?
 
  \vspace*{-0.5\baselineskip}
  \begin{tabular}{@{}ll@{}}
    \textit{False} & \textit{True} \\
    275886 & 11530
  \end{tabular}

  \item \textbf{Is on X-axis or Y-axis [XY]}: \textit{2-class}\\
  Is this column encoded on the X-axis or the Y-axis?
  
  \begin{tabular}{@{}ll@{}}
    \textit{False} & \textit{True} \\
    144364 & 142814
  \end{tabular}
 
\end{enumerate}
\end{mdframed}

For the \textbf{Visualization Type} and \textbf{Mark Type} tasks, the 2-class task predicts line vs. bar, and the 3-class predicts scatter vs. line vs. bar. Though Plotly supports over twenty mark types, we limited prediction outcomes to the few types that comprise the majority of visualizations within our corpus. This heterogeneity of visualization types is consistent with the findings of \cite{battle-beagle, public-data-and-visualizations}.

\begin{table*}[t!]
\begin{subtable}[t]{0.47\textwidth}
    \caption{Prediction accuracies for two visualization-level tasks.}
    \label{tab:task-accuracies-visualization-level}
    \begin{tabularx}{\textwidth}{@{}lll*4{>{\centering\arraybackslash}X}@{}}
    \toprule
     &  & \textbf{} & \multicolumn{3}{c}{\textbf{Visualization Type}} & \multicolumn{1}{c}{\textbf{HSA}} \\ 
     \cmidrule(l){4-7} 
    \textbf{Model} & \textbf{Features} & \textbf{d} & \multicolumn{1}{c}{\textit{C=2}} & \multicolumn{1}{c}{\textit{C=3}} & \multicolumn{1}{c}{\textit{C=6}} & \multicolumn{1}{c}{\textit{C=2}} \\ \midrule
    NN & D & \multicolumn{1}{l|}{15} & 66.3 & 50.4 & 51.3 & 84.1 \\
     & D+T & \multicolumn{1}{l|}{52} & 75.7 & 59.6 & 60.8 & 86.7 \\
     & D+T+V & \multicolumn{1}{l|}{717} & 84.5 & 77.2 & 87.7 & 95.4 \\
     & All & \multicolumn{1}{l|}{841} & \textbf{86.0} & \textbf{79.4} & \textbf{89.4} & \textbf{97.3} \\ \midrule
    NB & All & \multicolumn{1}{l|}{841} & 63.4 & 49.5 & 46.2 & 72.9 \\
    KNN & All & \multicolumn{1}{l|}{841} & 76.5 & 59.9 & 53.8 & 81.5 \\
    LR & All & \multicolumn{1}{l|}{841} & \textbf{81.8} & 64.9 & \textbf{69.0} & 90.2 \\ 
    RF & All & \multicolumn{1}{l|}{841} & 81.2 & \textbf{65.1} & 66.6 & \textbf{90.4} \\ \midrule
    \multicolumn{3}{l|}{$\mathbf{N_{raw}}$ (in 1000s)} & 42.2 & 87.0 & 99.3 & 119 \\ \bottomrule
    
    \end{tabularx}
\end{subtable}
\hfill
\begin{subtable}[t]{0.52\textwidth}
    \caption{Prediction accuracies for three encoding-level tasks.}
    \label{tab:task-accuracies-column-level}
    \begin{tabularx}{\textwidth}{@{}lll*5{>{\centering\arraybackslash}X}@{}}
    \toprule
     &  & \textbf{} & \multicolumn{3}{c}{\textbf{Mark Type}} & \multicolumn{1}{c}{\textbf{ISA}} & \multicolumn{1}{c}{\textbf{XY}} \\ \cmidrule(l){4-8} 
    \textbf{Model} & \textbf{Features} & \textbf{d} & \multicolumn{1}{c}{\textit{C=2}} & \multicolumn{1}{c}{\textit{C=3}} & \multicolumn{1}{c}{\textit{C=6}} & \multicolumn{1}{c}{\textit{C=2}} & \multicolumn{1}{c}{\textit{C=2}} \\ \midrule
    NN & D & \multicolumn{1}{l|}{1} & 65.2 & 44.3 & 30.5 & 52.1 & 49.9 \\
     & D+T & \multicolumn{1}{l|}{9} & 68.5 & 46.8 & 35.0 & 70.3 & 57.3 \\
     & D+T+V & \multicolumn{1}{l|}{66} & 79.4 & 59.4 & 76.0 & 95.5 & 67.4 \\
    & All & \multicolumn{1}{l|}{81} & \textbf{84.9} & \textbf{67.8} & \textbf{82.9} & \textbf{98.3} & \textbf{83.1} \\ \midrule
    NB & All & \multicolumn{1}{l|}{81} & 57.6 & 41.1 & 27.4 & 81.2 & 70.0 \\
    KNN & All & \multicolumn{1}{l|}{81} & 72.4 & 51.9 & 37.8 & 72.0 & 65.6 \\
    LR & All & \multicolumn{1}{l|}{81} & 73.6 & 52.6 & 43.7 & \textbf{84.8} & 79.1 \\ 
    RF & All & \multicolumn{1}{l|}{81} & \textbf{78.3} & \textbf{60.1} & \textbf{46.7} & 74.2 & \textbf{83.4} \\ \midrule
    \multicolumn{3}{l|}{$\mathbf{N_{raw}}$ (in 1000s)} & 94.7 & 163 & 183 & 287 & 287 \\ \bottomrule
    \end{tabularx}
\end{subtable}
\caption{Design choice prediction accuracies for five models, averaged over 5-fold cross-validation. The standard error of the mean was $<0.1\%$ for all results. Results are reported for a neural network (NN), naive Bayes (NB), K-nearest neighbors (KNN), logistic regression (LR), and random forest (RF). Features are separated into four categories: dimensions (D), types (T), values (V), and names (N). $\mathbf{N_{raw}}$ is the size of the training set before resampling, \textbf{d} is the number of features, \textit{C} is the number of outcome classes.  \textbf{HSA} = Has Shared Axis, \textbf{ISA} = Is Shared X-axis or Y-Axis and \textbf{XY} = Is on X-axis or Y-axis.}\label{tab:task-accuracies}

\end{table*}

\subsection{Neural Network and Baseline Models}

Our primary model is a fully-connected feedforward neural network (NN), which consists of non-linear functions connected as nodes in a network. Our network had $3$ hidden layers, each consisting of $1,000$ neurons with ReLU activation functions.

We chose four simpler models as baselines: naive Bayes (NB), which makes predictions based on conditional probabilities determining by applying Bayes' theorem while assuming independent features; K-nearest neighors (KNN), which predicts based on the majority vote of the $K$ most similar points; logistic regression (LR), a generalized linear model that predicts the probability of a binary event with a logistic function; and random forests (RF), an ensemble of decision trees that continually split the input by individual features.

We implemented the NN using PyTorch~\cite{pytorch}, and the baseline models using scikit-learn. The baseline models used default scikit-learn~\cite{scikit} parameters. Specifically, KNN used 5 neighbors with a Euclidean distance metric. LR used an L1 regularization penalization norm, and a regularization strength of $1$. RF had no maximum depth, used Gini impurity criteria, and considered $\sqrt{d}$ features when looking for a split, where $d$ is the total number of features. Randomized parameter search did not result in a significant performance increase over the results reported in the next section.

\subsection{Training and Testing Models}
The neural network was trained with the Adam optimizer and mini-batch size of $200$. The learning rate was initialized at $5\times10^{-4}$, and followed a learning rate schedule that reduces the learning rate by a factor of $10$ upon encountering a plateau. A plateau was defined as $10$ epochs with validation accuracy that do vary beyond a threshold of $10^{-3}$. Training ended after the third decrease in the learning rate, or at $100$ epochs. We found that weight decay and dropout did not significantly improve performances.  

For the neural network, we split the data into 60/20/20 train/validation/test sets. That is, we train the NN on 60\% of the data to optimize performance on a separate 20\% validation set. Then, we evaluate performance at predicting the remaining 20\% test set. For the baseline models, which do not require a validation set, we used a 60/20 train/test split. 

We oversample the train, validation, and test sets to the size of the majority class and ensure no overlap between the three sets. We oversample for two reasons. First, because of the heterogeneous outcomes, naive classifiers guessing the base rates would have high accuracies. Second, for ease of interpretation, balanced classes allow us to report standard accuracies, which is ideal for prediction tasks with number of outcome classes $C>2$. 

We train and test each model five times (5-fold cross-validation), so that each sample in the corpus was included in exactly one test set. We report the average performance across these tests. Reported results are the average of 5-fold cross-validation, such that each sample in total corpus was included in exactly one test set.

In terms of features, we constructed four different feature sets by incrementally adding the \textbf{Dimensions (D)},  \textbf{Types (T)}, \textbf{Values (V)}, and \textbf{Names (N)} categories of features, in that order. We refer to these feature sets as \textbf{D}, \textbf{D+T}, \textbf{D+T+V}, and \textbf{D+T+V+N=All}. The neural network was trained and tested using all four feature sets. The four baseline models only used the full feature set (D+T+V+N=All).

Lastly, we use accuracy (fraction of correct predictions) instead of other measures of performance, such as $F_1$ score and AUROC, because it easily generalizes to multi-class cases, its straight-forward interpretation, and because we equally weigh the outcomes of our tasks.
\section{Evaluating Prediction Performance}\label{sec:results}

We report performance of each model on the seven prediction tasks in Table~\ref{tab:task-accuracies}. The highest achieved mean accuracies for both the neural network and the baseline models are highlighted in bold. The top accuracies are achieved by the neural network. Across the board, each model achieved accuracies above the random guessing baseline of $(1/C)\%$ (\textit{e.g.} 50\% accuracy on the two-type visualization type prediction task). Model performance generally progressed as NB $<$ KNN $<$ LR $\approx$ RF $<$ NN. That said, the performance of both RF and LR is not significantly lower than that of the NN in most cases. Simpler classifiers may be desirable, depending on the need for optimized accuracy, and the trade-off with other factors such as interpretability and training cost.

\begin{figure}[t!]
    \centering
    \begin{subfigure}{\columnwidth}
        \includegraphics[width=\columnwidth]{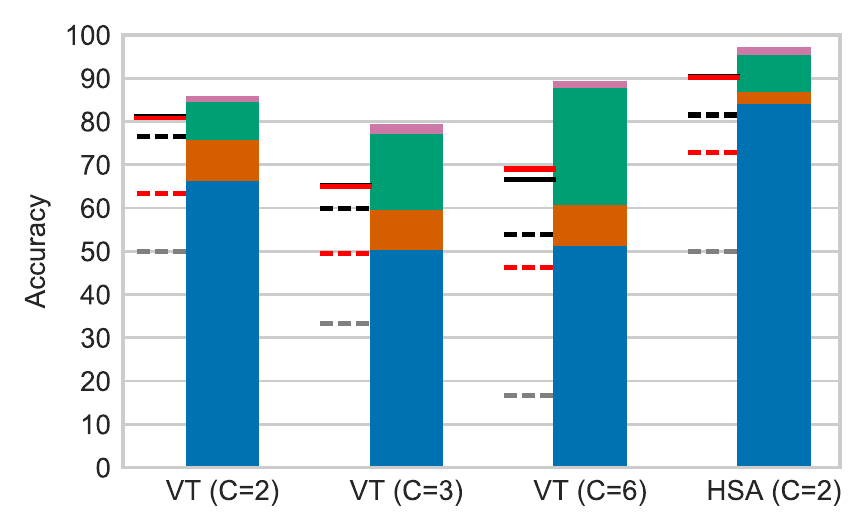}
        \caption{Marginal accuracies by feature set for visualization-level prediction tasks.}
        \label{fig:marginal-accuracy-chart-tasks}
    \end{subfigure}
    
    \vspace{1em}

    \begin{subfigure}{\columnwidth}
        \includegraphics[width=\columnwidth]{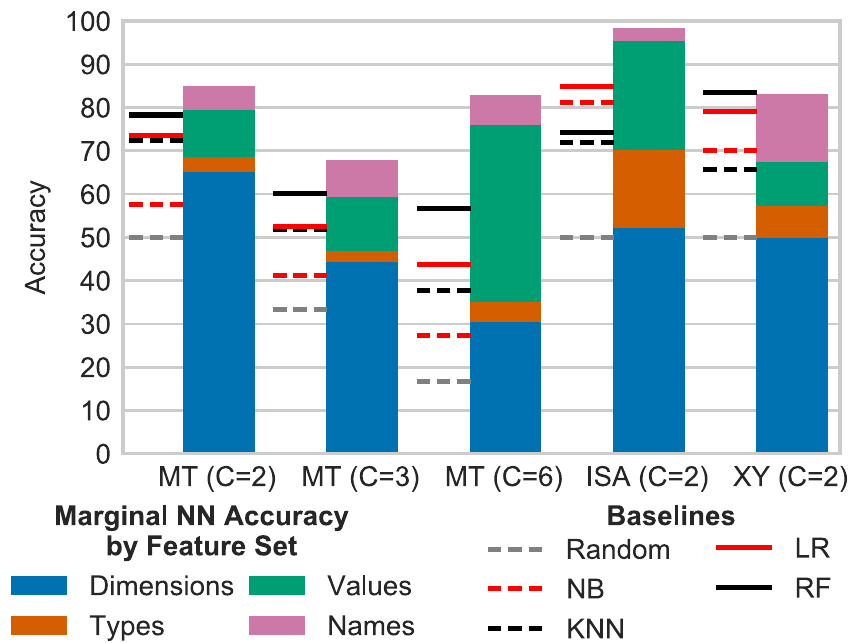}
        \caption{Marginal accuracies by feature set for encoding-level prediction tasks.}
        \label{fig:marginal-accuracy-trace-tasks}
    \end{subfigure}
    
    \caption{Marginal contribution to NN accuracy by feature set, for each task. Baseline accuracies are shown as solid and dashed lines.}\label{fig:marginal-accuracy}
\end{figure}

\begin{table*}[t!]
    \scriptsize
    \begin{subtable}{\textwidth}
        \caption{Feature importances for two visualization-level prediction tasks. The second column describes how each feature was aggregated, using the abbreviations in Table~\ref{tab:aggregations}.}
        \label{tab:feature-importances-visualization-level}
        \begin{tabularx}{\textwidth}{@{}r*4{lXl}@{}}
        \toprule
        \# & \multicolumn{3}{c}{\textbf{Visualization Type} \textit{(C=2)}} & \multicolumn{3}{c}{\textbf{Visualization Type} \textit{(C=3)}} & \multicolumn{3}{c}{\textbf{Visualization Type} \textit{(C=6)}} & \multicolumn{3}{c}{\textbf{Has Shared Axis} \textit{(C=2)}} \\ \midrule
        \multicolumn{1}{l|}{1} & \cellcolor[HTML]{F58518} & \% of Values are Mode & \multicolumn{1}{l|}{std} & \cellcolor[HTML]{F58518} & Entropy & \multicolumn{1}{l|}{std} & \cellcolor[HTML]{54A24B} & Is Monotonic & \multicolumn{1}{l|}{\%} & \cellcolor[HTML]{BAB0AC} & Number of Columns &  \\
        \multicolumn{1}{l|}{2} & \cellcolor[HTML]{F58518} & Min Value Length & \multicolumn{1}{l|}{max} & \cellcolor[HTML]{F58518} & Entropy & \multicolumn{1}{l|}{var} & \cellcolor[HTML]{BAB0AC} & Number of Columns & \multicolumn{1}{l|}{} & \cellcolor[HTML]{54A24B} & Is Monotonic & \% \\
        \multicolumn{1}{l|}{3} & \cellcolor[HTML]{F58518} & Entropy & \multicolumn{1}{l|}{var} & \cellcolor[HTML]{EECA3B} & String Type & \multicolumn{1}{l|}{\%} & \cellcolor[HTML]{54A24B} & Sortedness & \multicolumn{1}{l|}{max} & \cellcolor[HTML]{B279A2} & Field Name Length & AAD \\
        \multicolumn{1}{l|}{4} & \cellcolor[HTML]{F58518} & Entropy & \multicolumn{1}{l|}{std} & \cellcolor[HTML]{F58518} & Mean Value Length & \multicolumn{1}{l|}{var} & \cellcolor[HTML]{B279A2} & Y In Name & \multicolumn{1}{l|}{\#} & \cellcolor[HTML]{B279A2} & \# Words In Name & NR \\
        \multicolumn{1}{l|}{5} & \cellcolor[HTML]{EECA3B} & String Type & \multicolumn{1}{l|}{has} & \cellcolor[HTML]{F58518} & Min Value Length & \multicolumn{1}{l|}{var} & \cellcolor[HTML]{B279A2} & Y In Name & \multicolumn{1}{l|}{\%} & \cellcolor[HTML]{B279A2} & X In Name & \# \\
        \multicolumn{1}{l|}{6} & \cellcolor[HTML]{F58518} & Median Value Length & \multicolumn{1}{l|}{max} & \cellcolor[HTML]{EECA3B} & String Type & \multicolumn{1}{l|}{has} & \cellcolor[HTML]{FF9DA6} & \# Shared Unique Vals & \multicolumn{1}{l|}{std} & \cellcolor[HTML]{B279A2} & \# Words In Name & range \\
        \multicolumn{1}{l|}{7} & \cellcolor[HTML]{F58518} & Mean Value Length & \multicolumn{1}{l|}{AAD} & \cellcolor[HTML]{F58518} & Percentage Of Mode & \multicolumn{1}{l|}{std} & \cellcolor[HTML]{FF9DA6} & \# Shared Values & \multicolumn{1}{l|}{MAD} & \cellcolor[HTML]{B279A2} & Edit Distance & mean \\
        \multicolumn{1}{l|}{8} & \cellcolor[HTML]{F58518} & Entropy & \multicolumn{1}{l|}{mean} & \cellcolor[HTML]{F58518} & Median Value Length & \multicolumn{1}{l|}{max} & \cellcolor[HTML]{F58518} & Entropy & \multicolumn{1}{l|}{std} & \cellcolor[HTML]{B279A2} & Edit Distance & max \\
        \multicolumn{1}{l|}{9} & \cellcolor[HTML]{4C78A8} & Entropy & \multicolumn{1}{l|}{max} & \cellcolor[HTML]{F58518} & Entropy & \multicolumn{1}{l|}{mean} & \cellcolor[HTML]{F58518} & Entropy & \multicolumn{1}{l|}{range} & \cellcolor[HTML]{BAB0AC} & Length & std \\
        \multicolumn{1}{l|}{10} & \cellcolor[HTML]{F58518} & Min Value Length & \multicolumn{1}{l|}{AAD} & \cellcolor[HTML]{BAB0AC} & Length & \multicolumn{1}{l|}{mean} & \cellcolor[HTML]{F58518} & \% of Values are Mode & \multicolumn{1}{l|}{std} & \cellcolor[HTML]{B279A2} & Edit Distance & NR \\ \bottomrule
        \end{tabularx}
    \end{subtable}
    
    \begin{subtable}{\textwidth}
        \caption{Feature importances for four encoding-level prediction tasks.}
        \label{tab:feature-importances-encoding-level}
        \begin{tabularx}{\textwidth}{@{}r*5{lX}@{}}
        \toprule
        \# & \multicolumn{2}{c}{\textbf{Mark Type} \textit{(C=2)}} & \multicolumn{2}{c}{\textbf{Mark Type} \textit{(C=3)}} & \multicolumn{2}{c}{\textbf{Mark Type} \textit{(C=6)}} & \multicolumn{2}{c}{\textbf{Is Shared Axis} \textit{(C=2)}} & \multicolumn{2}{c}{\textbf{Is X or Y Axis} \textit{(C=2)}} \\ \midrule
        \multicolumn{1}{l|}{1} & \cellcolor[HTML]{4C78A8} & \multicolumn{1}{l|}{Entropy} & \cellcolor[HTML]{BAB0AC} & \multicolumn{1}{l|}{Length} & \cellcolor[HTML]{BAB0AC} & \multicolumn{1}{l|}{Length} & \cellcolor[HTML]{B279A2} & \multicolumn{1}{l|}{\# Words In Name} & \cellcolor[HTML]{B279A2} & Y In Name \\
        \multicolumn{1}{l|}{2} & \cellcolor[HTML]{BAB0AC} & \multicolumn{1}{l|}{Length} & \cellcolor[HTML]{4C78A8} & \multicolumn{1}{l|}{Entropy} & \cellcolor[HTML]{B279A2} & \multicolumn{1}{l|}{Field Name Length} & \cellcolor[HTML]{9D755D} & \multicolumn{1}{l|}{Unique Percent} & \cellcolor[HTML]{B279A2} & X In Name \\
        \multicolumn{1}{l|}{3} & \cellcolor[HTML]{54A24B} & \multicolumn{1}{l|}{Sortedness} & \cellcolor[HTML]{B279A2} & \multicolumn{1}{l|}{Field Name Length} & \cellcolor[HTML]{4C78A8} & \multicolumn{1}{l|}{Entropy} & \cellcolor[HTML]{B279A2} & \multicolumn{1}{l|}{Field Name Length} & \cellcolor[HTML]{B279A2} & Field Name Length \\
        \multicolumn{1}{l|}{4} & \cellcolor[HTML]{72B7B2} & \multicolumn{1}{l|}{\% Outliers (1.5IQR)} & \cellcolor[HTML]{54A24B} & \multicolumn{1}{l|}{Sortedness} & \cellcolor[HTML]{54A24B} & \multicolumn{1}{l|}{Sortedness} & \cellcolor[HTML]{54A24B} & \multicolumn{1}{l|}{Is Sorted} & \cellcolor[HTML]{54A24B} & Sortedness \\
        \multicolumn{1}{l|}{5} & \cellcolor[HTML]{B279A2} & \multicolumn{1}{l|}{Field Name Length} & \cellcolor[HTML]{E45756} & \multicolumn{1}{l|}{Lin Space Seq Coeff} & \cellcolor[HTML]{E45756} & \multicolumn{1}{l|}{Lin Space Seq Coeff} & \cellcolor[HTML]{54A24B} & \multicolumn{1}{l|}{Sortedness} & \cellcolor[HTML]{BAB0AC} & Length \\
        \multicolumn{1}{l|}{6} & \cellcolor[HTML]{E45756} & \multicolumn{1}{l|}{Lin Space Seq Coeff} & \cellcolor[HTML]{72B7B2} & \multicolumn{1}{l|}{\% Outliers (1.5IQR)} & \cellcolor[HTML]{4C78A8} & \multicolumn{1}{l|}{Kurtosis} & \cellcolor[HTML]{B279A2} & \multicolumn{1}{l|}{X In Name} & \cellcolor[HTML]{4C78A8} & Entropy \\
        \multicolumn{1}{l|}{7} & \cellcolor[HTML]{72B7B2} & \multicolumn{1}{l|}{\% Outliers (3IQR)} & \cellcolor[HTML]{4C78A8} & \multicolumn{1}{l|}{Gini} & \cellcolor[HTML]{4C78A8} & \multicolumn{1}{l|}{Gini} & \cellcolor[HTML]{B279A2} & \multicolumn{1}{l|}{Y In Name} & \cellcolor[HTML]{E45756} & Lin Space Seq Coeff \\
        \multicolumn{1}{l|}{8} & \cellcolor[HTML]{4C78A8} & \multicolumn{1}{l|}{Norm. Mean} & \cellcolor[HTML]{4C78A8} & \multicolumn{1}{l|}{Skewness} & \cellcolor[HTML]{4C78A8} & \multicolumn{1}{l|}{Normality Statistic} & \cellcolor[HTML]{E45756} & \multicolumn{1}{l|}{Lin Space Seq Coeff} & \cellcolor[HTML]{4C78A8} & \cellcolor[HTML]{FFFFFF}Kurtosis \\
        \multicolumn{1}{l|}{9} & \cellcolor[HTML]{4C78A8} & \multicolumn{1}{l|}{Skewness} & \cellcolor[HTML]{4C78A8} & \multicolumn{1}{l|}{Norm. Range} & \cellcolor[HTML]{4C78A8} & \multicolumn{1}{l|}{Norm Range} & \cellcolor[HTML]{4C78A8} & \multicolumn{1}{l|}{Min} & \cellcolor[HTML]{B279A2} & \# Uppercase Chars \\
        \multicolumn{1}{l|}{10} & \cellcolor[HTML]{4C78A8} & \multicolumn{1}{l|}{Norm. Range} & \cellcolor[HTML]{4C78A8} & \multicolumn{1}{l|}{Norm. Mean} & \cellcolor[HTML]{4C78A8} & \multicolumn{1}{l|}{Skewness} & \cellcolor[HTML]{BAB0AC} & \multicolumn{1}{l|}{Length} & \cellcolor[HTML]{4C78A8} & \multicolumn{1}{l|}{Skewness} \\ \bottomrule
        \end{tabularx}
    \end{subtable}

    \caption{Top-10 feature importances for chart- and encoding-level prediction tasks. Feature importance is determined by mean decrease impurity for the top performing random forest models. Colors represent different feature groupings: dimensions (\colorlegend{dimensions}), type (\colorlegend{type}), statistical [Q] (\colorlegend{statistical_q}), statistical [C] (\colorlegend{statistical_c}), sequence (\colorlegend{sequence}), scale of variation (\colorlegend{space}), outlier (\colorlegend{outlier}), unique (\colorlegend{unique}), name (\colorlegend{name}), and pairwise-relationship (\colorlegend{pairwise}).
    }
    \label{tab:feature-importances}
\end{table*}

Because the four feature sets are a sequence of supersets (D $\subset$ D+T $\subset$ D+T+V $\subset$ D+T+V+N), we consider the accuracy of each feature set above and beyond the previous. For instance, the increase in accuracy of a model trained on D+T+V over a model trained on D+T is a measure of the contribution of value-based (V) features. These marginal accuracies are visualized alongside baseline model accuracies in Fig.~\ref{fig:marginal-accuracy-chart-tasks}.

We note that the value-based features (\textit{e.g.} the statistical properties of a column) contribute more to performance than the type-based features (\textit{e.g.} whether a column is categorical), potentially because there are many more value-based features than type-based features. Or, because many value-based features are dependent on column type, there may be overlapping information between value- and type-based features.

\section{Interpreting Feature Importances}\label{sec:features}

We calculate feature importances to interpret our models, justify our feature extraction pipeline, and relate our features to prior literature. Feature importances can also be used to inform visualization design guidelines, derive rules for rule-based systems, and perform feature selection for more parsimonious models.

Here, we determine feature importances for our top performing random forest models using the standard mean decrease impurity (MDI) measure~\cite{random-forest-variable-importances, classification-and-regression-trees}. The top ten features by MDI are shown in Table~\ref{tab:feature-importances-visualization-level}. We choose this method for its interpretability and its stability across runs. The reported features are generally consistent with those calculated through filter-based methods such as mutual information, or wrapper-based methods like recursive feature elimination.

We first note the importance of \textbf{dimensionality} (\begin{tikzpicture}\fill[fill=dimensions] (11.1,5.5) rectangle ++(0.4,0.2);\end{tikzpicture}), like the length of columns (\textit{i.e.} the number of rows) or the number of columns. For example, the length of a column is the second most important feature for predicting whether that column is visualized in a line or a bar trace. The dependence of mark type on number of visual elements is consistent with heuristics like ``keep the total number of bars under 12" for showing individual differences in a bar chart~\cite{visually}, and not creating pie charts with more ``more than five to seven" slices~\cite{eager-eyes}. The dependence on number of columns is related to the heuristics described by Bertin~\cite{Bertin:1983:SG:1095597} and encoded in Show Me~\cite{show-me}.

Features related to \textbf{column type} (\colorlegend{type}) are consistently important for each prediction task. For example, the whether a dataset has a string column is the fifth most important feature for determining whether that dataset is visualized as a bar or a line chart. The dependence of visualization type choice on column data type is consistent with the type-dependency of the perceptual properties of visual encodings described by Mackinlay and Cleveland and McGill~\cite{automating-the-design, cleveland-mcgill-graphical-perception}.

\textbf{Statistical features} (quantitative: \begin{tikzpicture}\fill[fill=statistical_q] (11.1,5.5) rectangle ++(0.4,0.2);\end{tikzpicture}, categorical: \begin{tikzpicture}\fill[fill=statistical_c] (11.1,5.5) rectangle ++(0.4,0.2);\end{tikzpicture}) such as Gini, entropy, skewness and kurtosis are important across the board. The presence of these higher order moments is striking because lower-order moments such as mean and variance are low in importance. The importance of these moments highlight the potential importance of capturing high-level characteristics of distributional shape. These observations support the use of statistical properties in visualization recommendation, like in~\cite{rank-by-feature, autovis}, but also the use of higher-order properties such as skewness, kurtosis, and entropy in systems such as Foresight~\cite{foresight}, VizDeck~\cite{perry2013vizdeck}, and Draco~\cite{draco}.

\textbf{Measures of orderedness} (\begin{tikzpicture}\fill[fill=sequence] (11.1,5.5) rectangle ++(0.4,0.2);\end{tikzpicture}), specifically sortedness and monotonicity, are important for many tasks as well. Sortedness is defined as the element-wise correlation between the sorted and unsorted values of a column, that is $|\textit{corr}(X_{raw}, X_{sorted})|$, which lies in the range $[0, 1]$. Monotonicity is determined by strictly increasing or decreasing values in $X_{raw}$. The importance of these features could be due to pre-sorting of a dataset by the user, which may reveal which column is considered to be the independent or explanatory column, which is typically visualized along the X-axis. While intuitive, we have not seen orderedness factor into existing systems.

We also note the importance of the linear or logarithmic space sequence coefficients, which are heuristic-based features that roughly capture the \textbf{scale of variation} (\begin{tikzpicture}\fill[fill=space] (11.1,5.5) rectangle ++(0.4,0.2);\end{tikzpicture}). Specifically, the linear space sequence coefficient is determined by $std(Y) / mean(Y)$, where  $Y = \{X_{i} - X_{i-1}\}_{i=(1+1)..N}$ for the linear space sequence coefficient, and $Y = \{X_{i} / X_{i-1}\}_{i=(1+1)..N}$ for the logarithmic space sequence coefficient. A column ``is" linear or logarithmic if its coefficient $\le10^{-3}$. Both coefficients are important in all four selected encoding-level prediction tasks. We have not seen similar measures of scale used in prior systems.

In sum, the contribution of these features to determining an outcome can be intuitive. In this way, these feature importances are perhaps unremarkable. However, the ability to quantitatively interpret these feature importances could serve as validation for visualization heuristics. Furthermore, the diversity of features in this list suggests that rule-based recommender systems, many of which incorporate only type information (\textit{e.g.}~\cite{show-me, 2017-voyager2}), should expand the set of considered features. This is computationally feasible because most features extracted by our system can be determined by inexpensive linear operations. That said, it would still be difficult in rule-based systems to capture the non-linear dependencies of task outcomes on features, and the complex relationships between features. 

\newpage
\section{Benchmarking with Crowdsourced Effectiveness}\label{sec:ground-truth}
We expand our definition of effectiveness from a binary to a continuous function that can be determined through crowdsourced consensus. Then, we describe our experimental procedure for gathering visualization type evaluations from Mechanical Turk workers. We compare different predictors at predicting these evaluations using a consensus-based effectiveness score.

\subsection{Modeling and Measuring Effectiveness}
As discussed in \cref{sec:problem}, we model data visualization as a process of making a set of design choices $C=\{c\}$ that maximize an effectiveness criteria $\textit{Eff}$ that depends on dataset $d$, task, and context. In \cref{sec:results}, we predict these design choices by training a machine learning model on a corpus of dataset-design choice pairs $[(d, c_{d})]$. But because each dataset was visualized only once by each user, we consider the user choices $c_{d}$ to be effective, and each other choice as ineffective. That is, we consider effectiveness to be binary.

But prior research suggests that effectiveness is continuous. For example, Saket et al. use time and accuracy preference to measure task performance~\cite{task-based-effectiveness}, Borkin et al. use a normalized memorability score~\cite{memorability}, and Cleveland and McGill use absolute error rates to measure performance on elementary perceptual tasks~\cite{cleveland-mcgill-graphical-perception}. Discussions by visualization experts~\cite{storytelling-with-data, to-optimize-or-to-satisfice} also suggest that multiple visualizations can be equally effective at displaying the same data.

Our effectiveness metric should be continuous and reflect the ambiguous nature of data visualization, which leads to multiple choices receiving a non-zero or even maximal score for the same dataset. This is in agreement with measures of performance for other machine learning tasks such as the BLEU score in language translation~\cite{bleu} and the ROUGE metric in text summarization~\cite{rouge}, where multiple results can be partly correct.

To estimate this effectiveness function, we need to observe a dataset $d$ visualized by $U$ potential users: $[(d, c_{d,1}), \ldots, (d, c_{d,U})]$. Assume that a design choice $c$ can take on multiple discrete values $\{v\}$. For instance, we consider $c$ the choice of \textbf{Visualization Type}, which can take on the values $\{\textit{bar}, \textit{line}, \textit{scatter}\}$. Using $n_v$ to denote the number of times $v$ was chosen, we compute the probability of making choice $v$ as $\hat{P}_{c}(v)=n_v / N$, and use $\{\hat{P}_c\}$ to denote the collection of probabilities across all $v$. We normalize the probability of choice $v$ by the maximum probability to define an effectiveness score:

\begin{equation}
    \hat{\textit{Eff}_c}(v) =\frac{\hat{P}_c(v)}{\textit{max}(\{\hat{P}_c\})}
    \label{normalized-effectiveness}
\end{equation}

Now, if all $N$ users make the same choice $v$, only $c=v$ will get the maximimum score while every other choice $c \neq v$ will receive a zero score. However, if two choices are chosen with an equal probability and are thus both equally effective, the normalization will ensure that both receive a maximum score.

Developing this crowdsourced score that reflects the ambiguous nature of making data visualization choices serves three main purposes. First, it lets us establish uncertainty around our models -- in this case, by bootstrap. Second, it lets us test whether models trained on the Plotly corpus can generalize and if Plotly users are actually making optimal choices. Lastly, it lets us benchmark against performance of the Plotly users as well as other predictors.

\subsection{Data Preparation}

To select the datasets in our benchmarking test set, we first randomly surfaced a set of candidate datasets that were visualized as either a bar, line, or scatter chart. Then, we removed obviously incomplete visualizations (\textit{e.g.} blank visualizations). Finally, we removed datasets that could not be visually encoded in all three visualization types without losing information. From the remaining set of candidates, we randomly selected 33 bar charts, 33 line charts, and 33 scatter charts.

As we cleaned the data, we adhered to four principles: modify the user's selections as little as possible, apply changes consistently to every dataset, rely on Plotly defaults, and don't make any change that is not obvious. For each of these datasets, we modified the raw column names to remove Plotly-specific biases (\textit{e.g.} removing ``\texttt{,x}" or ``\texttt{,y}" that was automatically append to column names). We also wanted to make the user evaluation experience as close to the original chart creation experience as possible. Therefore, we changed column names from machine-generated types if they are obvious from the user visualization axis labels or legend (\textit{e.g.} the first column is unlabeled but visualized as \texttt{Sepal Width} on the X-axis). Because of these modifications, both the Plotly users and the Mechanical Turkers accessed more information than our model.

We visualized each of these 99 datasets as a bar, line, and scatter chart. We created these visualizations by forking the original Plotly visualization then modifying Mark Types using Plotly Chart Studio. We ensured that color choices and axis ranges were consistent between all visualization types. The rest of the layout was held constant to the user's original specification, or the defaults provided by Plotly.

\subsection{Crowdsourced Evaluation Procedure}

\begin{figure}[t!]
 \centering
\includegraphics[width=\columnwidth]{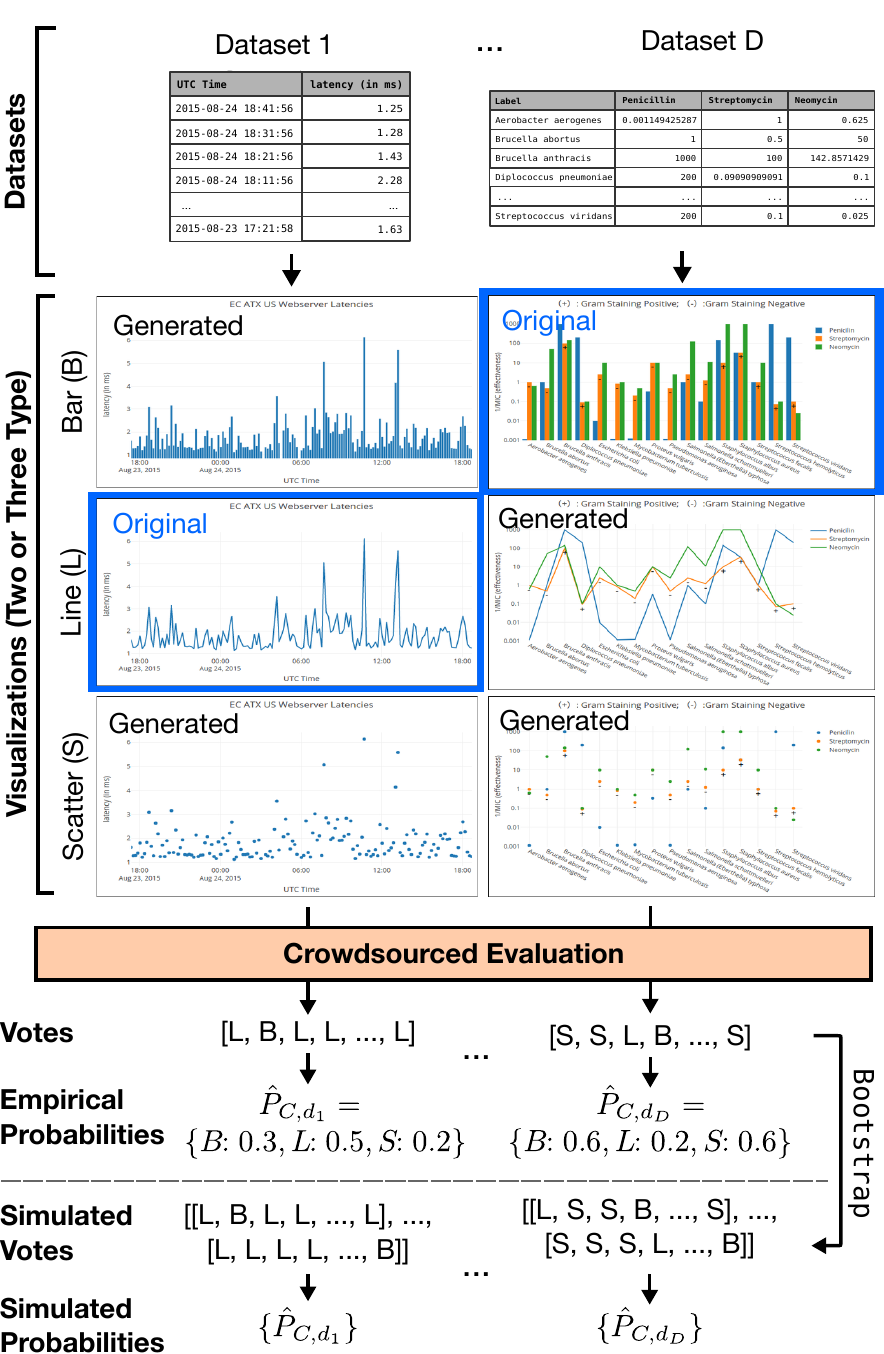}
 \caption{Experiment flow. The original user-generated visualizations are highlighted in blue, while we generated the visualizations of the remaining types. After crowdsourced evaluation, we have a set of votes for the best visualization type of that dataset. We calculate confidence intervals for model scores through bootstrapping.
 }
 \label{fig:experiment}
\end{figure}

We recruited participants through Amazon Mechanical Turk. To participate in the experiment, workers had to hold a U.S. bachelor degree and be at least 18 years of age, and be completing the survey on a phone. Workers also had to successfully answer three prescreen questions: 1) Have you ever seen a data visualization? [\textbf{Yes} or No], 2) Does the x-axis of a two-dimensional plot run horizontally or vertically? [\textbf{Horizontally}, Vertically, Both, Neither], 3) Which of the following visualizations is a bar chart? [\textbf{Picture of Bar Chart}, Picture of Line Chart, Picture of Scatter]. 150 workers successfully completed the two-class experiment, while 150 separate workers completed the three-class experiment. 

After successfully completing the pre-screen, workers evaluated the visualization type of 30 randomly selected datasets from our test set. Each evaluation had two stages. First, the user was presented the first 10 rows of the dataset, and told to "Please take a moment to examine the following dataset. (Showing first 10 out of X rows)." Then, after five seconds, the "next" button appeared. At the next stage, the user was asked "Which visualization best represents this dataset? (Showing first 10 out of X rows)." On this stage, the user was shown both the dataset and the corresponding bar, line, and scatter charts representing that dataset. A user could submit this question after a minimum of ten seconds. The evaluations were split into two groups of 15 by an attention check question. Therefore, each of the 66 datasets were evaluated $68.18$ times on average, while each of the $99$ ground truth datasets was evaluated $30$ times on average.  


\subsection{Benchmarking Procedure}
We use three types of predictors in our benchmark: human, model, and baseline. The two human predictors are the \textbf{Plotly} predictor, which is the visualization type of the original plot created by the Plotly user, and the \textbf{MTurk} predictor is the choice of a single random Mechanical Turk participant. When evaluating the performance of individual Mechanical Turkers, that individual's vote was excluded from the set of vote used in the mode estimation. 

The two learning-based predictors are \textbf{DeepEye} and \textbf{Data2Vis}. In both cases, we tried to make choices that maximize their CARS, within reason. We uploaded datasets to DeepEye as comma-separated values (CSV) files, and to Data2Vis as JSON objects. Unlike VizML and Data2Vis, DeepEye supports pie, bar, and scatter visualization types. We marked both pie and bar recommendations were both bar predictions, and scatter recommendations as line predictions in the two-type case. For both tools, we modified the data within reason to maximize the number of valid results. For the remaining errors (4 for Data2Vis and 14 for DeepEye), and cases without returned results (12 for DeepEye) we assigned a random chart prediction.

We evaluate the performance of a predictor using a score that assigns points to estimators based on the normalized effectiveness of a predicted value, from Equation~\ref{normalized-effectiveness}. This \textit{Consensus-Adjusted Recommendation Score} (CARS) of a predictor is defined as:

\begin{mdframed}[nobreak=true, backgroundcolor=black!10]
\begin{equation}\label{eq:cars}
    CARS_{predictor} = \frac{1}{|D|} \sum_{d \in D}{\frac{\hat{P}_c({\hat{c}_{predictor,\,d}})}{\textit{max}(\{\hat{P}_c\})}} \times 100
\end{equation}
\end{mdframed}

where $|D|$ is the number of datasets ($66$ for two-class and $99$ for three-class), $\hat{c}_{predictor,\,d}$ is the predicted visualization type for dataset $d$, and $\hat{P}_{c}$ returns the fraction of Mechanical Turker votes for a given visualization type. Note that the minimum CARS $>0\%$. We establish 95\% confidence intervals around these scores by comparing against $10^5$ bootstrap samples of the votes, which can be thought of as synthetic votes drawn from the observed probability distribution.

\subsection{Benchmarking Results}
\begin{figure}[t!]
    \centering
    \includegraphics[width=\columnwidth]{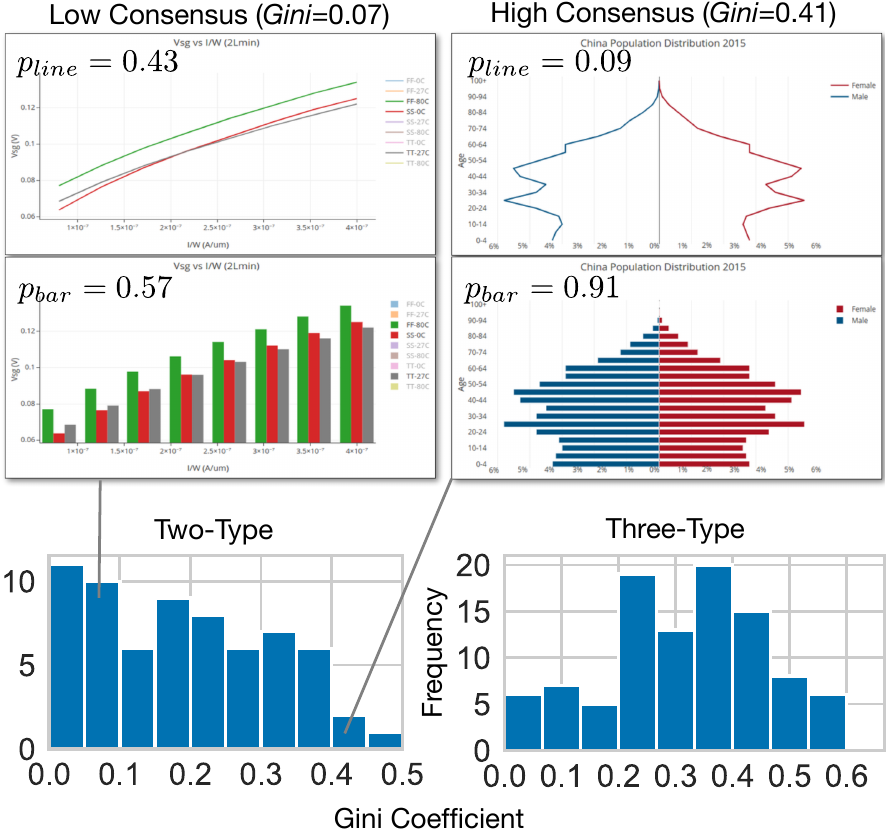}
    \caption{Distribution of Gini coefficients}  
    \label{fig:two-and-three-type-Gini}
\end{figure}

We first measure the degree of consensus using the Gini coefficient, the distribution of which is shown in Fig.~\ref{fig:two-and-three-type-Gini}. If a strong consensus was reached for all visualizations, then the Gini distributions would be strongly skewed towards the maximum, which is $1/2$ for the two-class case, and $2/3$ for the three-class case. Conversely, a lower Gini implies a weaker consensus, indicating an ambiguous ideal visualization type. The Gini distributions are not skewed towards either extreme, which supports the use of a soft scoring metric such as CARS over a hard measure like accuracy.

\begin{figure}[t!]
    \begin{subfigure}{\columnwidth}
        \includegraphics[width=\columnwidth]{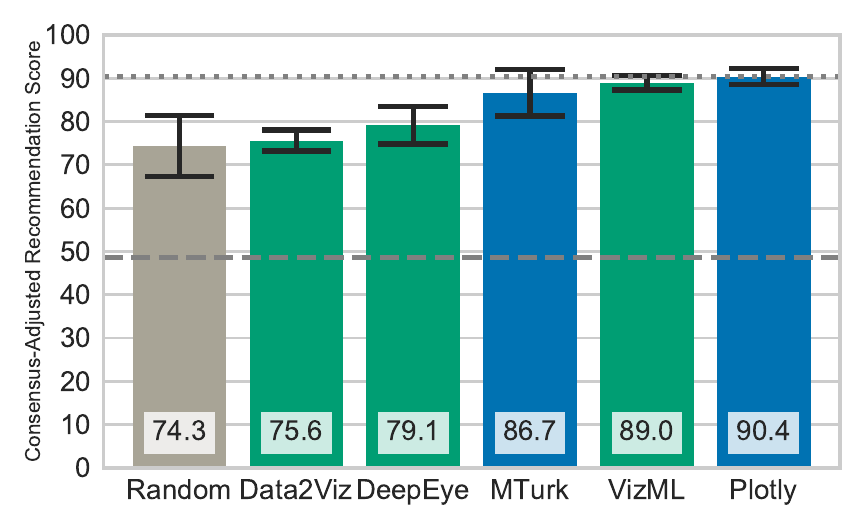}
        \vspace{-15pt}
        \caption{Two-type (bar vs. line) visualization type CARS.}
        \label{fig:two-type-experiment}
    \end{subfigure}

    \begin{subfigure}{\columnwidth}
        \includegraphics[width=\columnwidth]{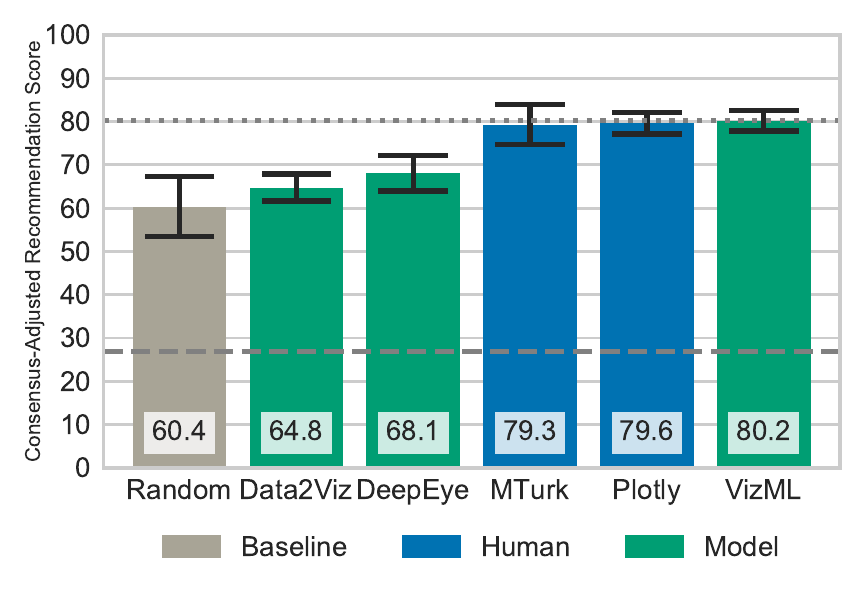}
        \vspace{-20pt}
        \caption{Three-type (bar vs. line vs. scatter) visualization type CARS.}
        \label{fig:three-type-experiment}
    \end{subfigure}
    \caption{Consensus-Adjusted Recommendation Score of three ML-based and two human predictors when predicting consensus visualization type. Error bars show 95\% bootstrapped confidence intervals, with $10^5$ bootstraps. The mean minimum achievable score is the lower dashed line, while the highest achieved CARS is the upper dotted line.}
    \label{fig:cars-bar-chart}
\end{figure}

The Consensus-Adjusted Recommendation Scores for each model and task are visualized as a bar chart in Fig.~\ref{fig:cars-bar-chart}. We first compare the CARS of VizML ($88.96\pm1.66$) against that of Mechanical Turkers ($86.66\pm5.38$) and Plotly users ($90.35\pm1.85$) for the two-class case, as shown in Fig.~\ref{fig:two-type-experiment}. It is surprising that VizML performs comparably to the original Plotly users, who possess domain knowledge and invested time into visualizing their own data. VizML significantly out-performs Data2Vis ($75.61\pm2.44$) and DeepEye ($79.12\pm4.33$). While neither Data2Vis nor DeepEye were trained to perform visualization type prediction, it is promising for ML-based recommender systems that both perform slightly better than the random classifier ($74.30\pm7.09$). For this task, the absolute minimum score was ($48.61\pm2.95$).

The same results are true for the three-class case shown in Fig.~\ref{fig:three-type-experiment}, in which the CARS of VizML ($81.18\pm2.39$) is slightly higher, but within error bars, than that of Mechanical Turkers ($79.28\pm4.66$), and Plotly users ($79.58\pm2.44$). Data2Vis ($64.75\pm3.13$) and DeepEye ($68.09\pm4.11$) outperform the Random ($60.37\pm6.98$) with a larger margin, but still within error. The minimum score was ($26.93\pm3.46$).

\section{Discussion}

In this paper, we introduce VizML, a machine learning approach to visualization recommendation using a large corpus of datasets and corresponding visualizations. We identify five key prediction tasks and show that neural network classifiers attain high test accuracies on these tasks, relative to both random guessing and simpler classifiers. We also benchmark with a test set established through crowdsourced consensus, and show that the performance of neural networks is comparable that of individual humans.

We acknowledge the limitations of this corpus and our approach. First, despite aggressive deduplication, our model is certainly biased towards the Plotly dataset. This bias could manifest on the user level (Plotly draws certain types of analysts), the system level (Plotly encourages or discourages certain types of plots, either by interface design or defaults), or the dataset level (Plotly is appropriate only for smaller datasets). We discuss approaches to improving the generalizability of VizML in the next section. 

Second, neither the Plotly user nor the Mechanical Turker is an expert in data visualization. However, if we consider laypeople the target audience of visualizations, the consensus opinion of crowdsourced agents may be a good measure of visualization quality.
Thirdly, we acknowledge that this paper was only focused on a subset of the tasks usually considered in a visualization recommendation pipeline. An ideal user-facing tool would include functionality that supports all tasks in the pipeline.

Yet, the high within-corpus test accuracies, and performance on the consensus dataset comparable to that of humans, lead us to claim that the structural and statistical properties of datasets influence how they are visualized. Furthermore, machine learning, by virtue of the ability to use or learn complex features for many datasets, can take advantage of these properties to augment the data visualization process. 

Machine learning tasks like image annotation or medical diagnosis are often objective, in that there exists a clear human-annotated ground truth. Other tasks are subjective, like language translation or text summarization tasks, which are benchmarked by human evaluation or against human-generated results. The question remains: is data visualization an objective or subjective process? Because of the high accuracies, we claim that there are definite regularities in how humans choose to visualize data that can be captured and leveraged by machine learning models. However, because crowd-sourced agents themselves do not agree with the consensus all of the time, there is an element of subjectivity in making visualization design choices.

\section{Future Research Directions}\label{sec:future-work}
To close, we discuss promising directions towards improving the data, methods, and tasks of machine learning-based recommender systems.

\paragraph{Public Training and Benchmarking Corpuses}
Despite the increasing prevalence of recommendation features within visualization tools, research progress in visualization recommendation is impeded due to the lack of a standard benchmark. Without a benchmark, it is difficult to bootstrap a recommender system or compare different approaches to this problem. Just as large repositories like ImageNet~\cite{imagenet_cvpr09} and CIFAR-10 played a significant role in shaping computer vision research, and serve as a useful benchmarkings, the same should exist for visualization recommendation.

\paragraph{Diverse Data Sources} By using Plotly data, we constrain ourselves to the final step of data visualization by assuming that the datasets are clean, and that a visualization encodes all columns of data. Yet, ``upstream" tasks like feature selection and data transformation are some of the most time-consuming tasks in data analysis. Tools like Tableau and Excel, which support selective visualization of columns and data transformation, could potentially provide the data needed to train models to augment these tasks.

\paragraph{Transfer Learning}
One explanation for the lack of prior ML-based visualization recommendation systems is the lack of available training data. Though our approach with using public data increases the size of the training set by an order of magnitude relative to that used by other systems, the monotonically increasing (unsaturated) learning curves of our models suggest that there is still room for more data. A common approach in other machine learning applications is to employ \textit{transfer learning}~\cite{transfer-learning}, which uses models trained on one task to scaffold a model on another task. For example, just as many neural networks in computer vision are pretrained on ImageNet, visualization recommendation models can be pretrained on the Plotly corpus and then transferred to domains with smaller training corpus sizes.

\paragraph{Representation Learning}
An approach trained on features extracted from raw data lends itself to straightforward interpretation and the use of standard machine learning models. But a representation learning approach trained on the raw data, instead of extracted features, has two advantages. First, it bypasses the laborious process of feature engineering. Second, via the universal approximation theorem for neural networks, it would be able to derive all hand-engineered features, and more, if important for predicting the outcome.

\paragraph{Unsupervised Learning}
Another approach to end-to-end visualization recommendation is to use semantic measures between datasets in a ``dataset space" with a traditional recommendation system (\textit{e.g.} model-based collaborative filtering). Initial explorations of unsupervised clustering techniques like t-distributed stochastic neighbor embedding (t-SNE)~\cite{t-sne} and UMAP suggest non-trivial structure in the dataset space.

\paragraph{Addressing the Multiple Comparisons Problem}
Analysts continually using visualization to both explore and confirm hypotheses are at risk of arriving at spurious insights, via the multiple comparisons problem (MCP)~\cite{multiple-comparisons-problem}. But if visual analytics tools are fishing rods for spurious insights, then visualizations recommender systems are deep ocean bottom trawlers. The MCP is exacerbated by opaque ML-based recommender systems, in which the number of implicit comparisons is difficult to track.

\paragraph{Integrating Prediction Tasks into Pipeline Model}
The ``holy grail" of visualization recommendation remains an end-to-end model which accepts a dataset as input and produces visualizations as output, which can then be evaluated in a user-facing system. An end-to-end model based on our approach of recommending design choices would combine the outcomes of each prediction task into a ranked list of recommendations. However, the predicted outcomes are sometimes inconsistent. The simplest approach is combining outcomes with heuristics. Two other approaches are generating a multi-task learning model that outputs all design choices, or to developing a pipeline model that predicts outcomes in sequence.

\acknowledgments{
The authors thank Owais Khan, \c{C}a\v{g}atay Demiralp, Sharon Zhang, Diana Orghian, Madelon Hulsebos, Laura Pang, David Alvarez-Melis, and Tommi Jaakkola for their feedback. We also thank Alex Johnson and Plotly for making the  Community Feed data available. This work was supported in part by the MIT Media Lab consortium.
}

\newpage

\begin{appendices}
\section{Features and Aggregations}~\label{sec:features-and-aggregations}
\begin{table}[h!]
    \scriptsize
    
    \begin{subtable}[t]{\textwidth}
        \caption{\textbf{81 single-column features} describing the dimensions, types, values, and names of individual columns.}
        \label{tab:features-list_single-column}
        \begin{tabularx}{\columnwidth}{@{}p{1.9cm}X@{}}
            \hline
            \multicolumn{2}{c}{\cellcolor{black!10}\textbf{Dimensions (1)}} \\ \hline
            \begin{tabular}[t]{@{}l@{}}Length (1)\end{tabular} & Number of values \\ \hline

            \multicolumn{2}{c}{\cellcolor{black!10}\textbf{Types (8)}} \\ \hline
            \begin{tabular}[t]{@{}l@{}}General (3)\end{tabular} & Categorical (C), quantitative (Q), temporal (T) \\
            Specific (5) & String, boolean, integer, decimal, datetime \\ \hline

            \multicolumn{2}{c}{\cellcolor{black!10}\textbf{Values (58)}} \\ \hline

            \begin{tabular}[t]{@{}l@{}}Statistical {[}Q, T{]}\\ (16)\end{tabular} & \begin{tabular}[t]{@{}l@{}}Mean, median, range × (Raw/normalized by max),\\ variance, standard deviation, coefficient of variance,\\ minimum, maximum, (25th/75th) percentile,\\ median absolute deviation, average absolute\\deviation, quantitative coefficient of dispersion\end{tabular} \\
            
            \begin{tabular}[t]{@{}l@{}}Distribution {[}Q{]}\\(14)\end{tabular} & \begin{tabular}[t]{@{}l@{}}Entropy,  Gini,  skewness,  kurtosis,  moments\\ (5-10),  normality  (statistic,  p-value), \\ is normal at (p $<$ 0.05, p $<$ 0.01).\end{tabular} \\
            
            \begin{tabular}[t]{@{}l@{}}Outliers (8)\end{tabular} & (Has/\%) outliers at (1.5 $\times$ IQR, 3 $\times$ IQR, 99\%ile, 3$\sigma$) \\
        
            \begin{tabular}[t]{@{}l@{}}Statistical {[}C{]} (7)\end{tabular} & \begin{tabular}[t]{@{}l@{}}Entropy, (mean/median) value length, (min, std,\\ max) length of values, \% of mode\end{tabular} \\
            
            \begin{tabular}[t]{@{}l@{}}Sequence (7)\end{tabular} & \begin{tabular}[t]{@{}l@{}}Is sorted, is monotonic, sortedness, (linear/log)\\space sequence coefficient, is (linear/space) space\end{tabular}\\
            Unique (3)& (Is/\#/\%) unique \\
            Missing (3) & (Has/\#/\%) missing values \\ \hline
            
            \multicolumn{2}{c}{\cellcolor{black!10}\textbf{Names (14)}} \\ \hline
            \begin{tabular}[t]{@{}l@{}}Properties (4)\end{tabular} & \begin{tabular}[t]{@{}l@{}}Name length, \# words, \# uppercase characters,\\starts with uppercase letter\end{tabular}\\
            
            \begin{tabular}[t]{@{}l@{}}Value (10)\end{tabular} & \begin{tabular}[t]{@{}l@{}}(``x", ``y", ``id", ``time", digit, whitespace, ``\textdollar",\\``\texteuro", ``\pounds", ``\textyen") in name\end{tabular} \\
            
            \bottomrule
        \end{tabularx}
    \end{subtable}

    \vspace{1em}
    \begin{subtable}[t]{\columnwidth}
        \caption{\textbf{30 pairwise-column features} describing the relationship between values and names of pairs of columns.}
        \label{tab:features-list_pairwise-column}
        \begin{tabularx}{\columnwidth}{@{}p{1.9cm}X@{}}
            \hline
            \multicolumn{2}{c}{\cellcolor{black!10}\textbf{Values (25)}} \\ \hline
            
            \begin{tabular}[t]{@{}l@{}}{[}Q-Q{]} (8)\end{tabular} & \begin{tabular}[t]{@{}l@{}}Correlation (value, $p$, $p<0.05$),\\Kolmogorov-Smirnov (value, $p$, $p<0.05$),\\(has, \%) overlapping range\end{tabular}\\
            
            \begin{tabular}[t]{@{}l@{}}{[}C-C{]} (6)\end{tabular} & \begin{tabular}[t]{@{}l@{}}$\chi^2$ (value, $p$, $p<0.05$),\\nestedness (value, $=1$, $>0.95\%$)\end{tabular}\\
            
            \begin{tabular}[t]{@{}l@{}}{[}C-Q{]} (3)\end{tabular} & \begin{tabular}[t]{@{}l@{}}One-Way ANOVA (value, $p$, $p<0.05$)\end{tabular}\\
            
            \begin{tabular}[t]{@{}l@{}}Shared values (8)\end{tabular} & \begin{tabular}[t]{@{}l@{}}is identical, (has/\#/\%) shared values, unique values\\are identical, (has/\#/\%) shared unique values\end{tabular}\\ \hline
            
            \multicolumn{2}{c}{\cellcolor{black!10}\textbf{Names (5)}} \\ \hline
            
            \begin{tabular}[t]{@{}l@{}}Character (2)\end{tabular} & \begin{tabular}[t]{@{}l@{}}Edit distance (raw/normalized)\end{tabular}\\
            
            \begin{tabular}[t]{@{}l@{}}Word (3)\end{tabular} & \begin{tabular}[t]{@{}l@{}}(Has, \#, \%) shared words\end{tabular}\\
            
            \bottomrule
        \end{tabularx}
    \end{subtable}

    \vspace{1em}
    \begin{subtable}[t]{\columnwidth}
        \caption{\textbf{16 Aggregation functions} used to aggregate single- and pairwise-column features into \textbf{841 dataset-level features}.}
        \label{tab:aggregations}
        \begin{tabularx}{\columnwidth}{@{}p{1.9cm}X@{}}
            \hline
            
            \begin{tabular}[t]{@{}l@{}}Categorical (5)\end{tabular} & \begin{tabular}[t]{@{}l@{}}Number (\#), percent (\%), has, only one (\#=1), all\end{tabular}\\
            
            \begin{tabular}[t]{@{}l@{}}Quantitative (10)\end{tabular} & \begin{tabular}[t]{@{}l@{}}Mean, variance, standard deviation, coefficient\\of variance (CV), min, max, range, normalized\\range (NR), average absolute deviation (AAD)\\ median absolute deviation (MAD)\end{tabular}\\
        
            \begin{tabular}[t]{@{}l@{}}Special (1)\end{tabular} & \begin{tabular}[t]{@{}l@{}}Entropy of data types\end{tabular}\\
            
            \bottomrule
        \end{tabularx}
    \end{subtable}
    
    \caption{Features and aggregation functions.}
    \label{tab:features-and-aggregations}
\end{table}

\end{appendices}

\bibliographystyle{abbrv}
\bibliography{references}

\end{document}